\documentstyle[prd,aps]{revtex}
\begin{document}
\input epsf
\draft
\renewcommand{\topfraction}{0.8}
\twocolumn[\hsize\textwidth\columnwidth\hsize\csname
@twocolumnfalse\endcsname
\preprint{SU-ITP-02-25, hep-ph/0205259, May 24, 2002}
\title { \Large   \bf Inflationary Theory versus Ekpyrotic/Cyclic Scenario}
\author{
Andrei Linde}
\address{ Department of Physics, Stanford University, Stanford CA
94305-4060, USA}
\date {May 24, 2002}
\maketitle
\begin{abstract}
I will discuss the development of inflationary theory and its present
status, as well as some recent attempts to suggest an
alternative to inflation. In particular, I will argue that the
ekpyrotic scenario in its original form does not solve any of the
major cosmological problems. Meanwhile, the cyclic scenario is not an
alternative to inflation but rather a complicated version of
inflationary theory. This scenario does not solve the flatness and
entropy problems, and it suffers from the singularity problem.  We
describe many other problems that need to be resolved in order to
realize a cyclic regime in this scenario, produce density
perturbations of a desirable magnitude, and preserve them after the
singularity. We propose several modifications of this scenario and
conclude that the best way to improve it is to add a usual stage of
inflation after the singularity and use that inflationary stage to
generate perturbations in the standard way. This modification
significantly simplifies the cyclic scenario, eliminates all of its
numerous problems, and makes it equivalent to the usual chaotic
inflation scenario.
\end{abstract}
\pacs{~~~~~{\it A talk at Stephen Hawking's $60^{th}$
 birthday conference, Cambridge University, Jan. 2002}}
 \vskip2pc]

\newcommand{\beq}{\begin{equation}}
\newcommand{\eeq}{\end{equation}}

\tableofcontents{}

\section{Introduction}

My first encounter with Stephen Hawking was related to inflationary
theory. It was quite dramatic. In the middle of October 1981 there was
a conference on Quantum Gravity in Moscow. This was the first
conference where I gave a talk on the new inflation scenario
\cite{New}. After my talk many participants of the conference from the
USA and Europe came up to me, asked questions, and even suggested
smuggling my paper abroad to speed up its publication. (The paper was
written in July 1981, but in accordance with Russian rules I spent 3 months
getting permission for its publication.)

Somehow I did not have a chance to discuss it with Stephen at the
conference, but we did it the next day, under rather unusual
circumstances. He was invited to give a talk at the Sternberg
Astronomy Institute of Moscow State University. His talk, based on his
work with Moss and Stewart \cite{HMS}, was about the problems of the
old inflationary theory proposed by Alan Guth \cite{Guth}.  The main conclusion of their work \cite{HMS}, as well as of the subsequent paper by Guth and Weinberg \cite{GuthWeinb}, was that it is impossible to improve the old inflation scenario.

 Rather unexpectedly, I was asked to translate. At that time Stephen did not
have his computer, so his talks usually were given by his students. He
would just sit around and add brief comments if a student would say
something wrong. This time, however, they were not quite
prepared. Stephen would say one word, his student would say one word,
and then I would translate this word, so in the beginning the talk
progressed very slowly.  Since I knew the subject, I started adding
lengthy explanations in Russian. Thus, Stephen would say one word, his
student would say one word, and then I would talk for few minutes.
Then Stephen would talk again, etc. Everything went smoothly during
the first part of the talk when we were explaining the problems of the
old inflationary theory.

Then Stephen said that recently Andrei Linde had suggested an interesting 
way to solve the problems of inflationary theory. I
happily translated this. The best physicists of Russia are here to
listen to Stephen, my future depends on them, and now he is going to
explain my work to them; what could be better? But then Stephen said
that the new inflationary scenario cannot work... and I
translated. For the next half hour I was translating for Stephen and
explaining to everyone the problems with my scenario and why it does
not work...

I do not remember ever being in any other situation like that. What
shall I do, what shall I do?.. When the talk was over I said that I
translated but I disagree, and explained why. Then I suggested to
Stephen that we discuss it privately. We found an empty office and for
almost two hours the authorities of the Institute were in a panic
searching for the famous British scientist who had miraculously
disappeared.  Meanwhile I was talking to him about various parts of
the new inflationary scenario. From time to time Stephen would say
something and his student would translate: ``But you did not say that
before.'' This was repeated over and over again. Then Stephen invited
me to his hotel where we continued the discussion. Then he started
showing me photos of his family and invited me to Cambridge. This was
the beginning of a beautiful friendship.

After that event the story developed at a rapid pace. In October I
sent my paper to Physics Letters and I also sent my preprints to many
places in the USA. After returning to England Stephen started working
on new inflation together with Ian Moss \cite{HawkMoss}. Three months later,
Paul Steinhardt and Andy Albrecht wrote a paper on new inflation with results
very similar to mine \cite{AlStein}. In Summer 1982 Stephen organized
a workshop in Cambridge dedicated to new inflation.  This was the best
and most productive workshop I have ever attended.

In a certain sense, this was the first and the last workshop on new
inflation. The theory of inflationary perturbations of scalar fields
\cite{Vilenkin:wt,Linde:uu}, as well as the theory of
post-inflationary density perturbations \cite{Mukh,Hawk,Mukh2}, were
to a large extent developed at this workshop \cite{Hawk}. Calculations
using these theories showed that the coupling constant of the scalar
field in new inflation had to be smaller than $10^{-12}$. Such a field
could not be in a state of thermal equilibrium in the early
universe. This means, in particular, that the theory of
high-temperature phase transitions \cite{Kirzhnits}, which served as
the basis for old and new inflation, was in fact irrelevant for
inflationary cosmology. Thus, some other approach was necessary.  The
assumption of thermal equilibrium requires many particles interacting
with each other. This means that new inflation could explain why our
universe was so large only if it was very large from the
beginning. Finally, inflation in this theory begins very late, and
during the preceding epoch the universe could easily have collapsed or
become so inhomogeneous that inflation could never happen.  In
addition, this scenario could work only if the effective potential of
the field $\phi$ had a very flat plateau near $\phi = 0$, which is
somewhat artificial. Because of all of these difficulties, no
realistic versions of the new inflationary universe scenario have been
proposed so far.

From a more general perspective, old and new inflation represented a
substantial but incomplete modification of the big bang theory. It was
still assumed that the universe was in a state of thermal equilibrium
from the very beginning, that it was relatively homogeneous and large
enough to survive until the beginning of inflation, and that the stage
of inflation was just an intermediate stage of the evolution of the
hot universe. In the beginning of the 80's these assumptions seemed
natural and practically unavoidable. That is why it was so difficult
to overcome a certain psychological barrier and abandon all of these
assumptions. This was done with the invention of the chaotic inflation
scenario~\cite{Chaot}. This scenario resolved all the problems
mentioned above. According to this scenario, inflation can occur even
in theories with the simplest potentials such as $V(\phi) \sim
\phi^n$.  Inflation can begin even if there was no thermal equilibrium
in the early universe, and it can even start close to the Planck
density, in which case the problem of initial conditions for inflation
can be easily resolved~\cite{book}.

Stephen was the first person (apart from my Russian colleagues) to
whom I spoke about chaotic inflation. Since that time in his work on
inflation he has used only this model, as well as some modifications
of the Starobinsky scenario~\cite{Star}. Let me describe the basic
features of chaotic inflation.

\section{Chaotic inflation}

Consider  the simplest model of a scalar field $\phi$ with a mass $m$ and with the
potential energy density $V(\phi)  = {m^2\over 2} \phi^2$.
Since this function has a minimum at $\phi = 0$,  one may expect that the
scalar field $\phi$ should oscillate near this minimum. This is indeed
the case if the universe does not expand, in which case equation of motion for the scalar field coincides with equation for harmonic oscillator, $\ddot\phi = -m^2\phi$.

However, because of the expansion of the universe with Hubble constant $H = \dot a/a $, an additional term $3H\dot\phi$ appears in the harmonic oscillator equation:
\begin{equation}\label{1}
 \ddot\phi + 3H\dot\phi = -m^2\phi \ .
\end{equation}
The term $3H\dot\phi$ can be interpreted as a friction term.
The Einstein equation for a homogeneous universe containing scalar field $\phi$ looks as follows:
\begin{equation}\label{2}
H^2 +{k\over a^2} ={1\over 6}\, \left(\dot \phi
^2+m^2 \phi^2) \right) \ .
\end{equation}
Here $k = -1, 0, 1$ for an open, flat or closed universe
respectively. We work in units $M_p^{-2} = 8\pi G = 1$. 

If   the scalar field $\phi$  initially was large,   the Hubble parameter $H$
was large too, according to the second equation. This means that the
friction term $3H\dot\phi$ was very large, and therefore    the
scalar field was moving   very slowly, as a ball in a viscous liquid.
Therefore at this stage the energy density of the scalar field, unlike
the  density of ordinary matter,   remained almost constant, and
expansion of the universe continued with a much greater speed than in the
old cosmological theory. Due to the rapid growth of the scale of the
universe and a slow motion of the field $\phi$, soon after the beginning
of this regime one has $\ddot\phi \ll 3H\dot\phi$, $H^2 \gg {k\over
a^2}$, $ \dot \phi ^2\ll m^2\phi^2$, so  the system of equations can be
simplified:
\begin{equation}\label{E04}
H= {\dot a \over a}   ={
m\phi\over \sqrt 6}\ , ~~~~~~  \dot\phi = -m\  \sqrt{2\over 3}     .
\end{equation}
The first equation shows that if the field $\phi$ changes slowly, the size of the universe in this regime
grows approximately as $e^{Ht}$, where $H = {m\phi\over\sqrt 6}$. This is the stage of inflation, which ends when the field $\phi$ becomes much smaller than $M_p=1$. Solution of these equations shows that after a long stage of inflation  the universe initially filled with the field $\phi = \phi_0 \gg 1$  grows  exponentially \cite{book}, 
\begin{equation}\label{E05}
a= a_0 \ e^{\phi_0^2/4}  .
\end{equation}

This is as simple as it could be. Inflation does not require supercooling and tunneling from the false vacuum \cite{Guth}, or rolling from an artificially flat top of the effective potential \cite{New,AlStein}. It appears in the theories that can be as simple as a theory of harmonic oscillator \cite{Chaot}. Only after I realized it, I started to believe that inflation is not a trick necessary to fix  problems of the old big bang theory, but a generic cosmological regime.

 In  realistic versions of inflationary theory the  duration of inflation could be as short as $10^{-35}$ seconds. When inflation ends, the
scalar field $\phi$ begins to   oscillate near the minimum of $V(\phi)$.
As any rapidly oscillating classical field, it looses its energy by
creating pairs of elementary particles. These particles interact with
each other and come to a state of thermal equilibrium with some
temperature $T$ \cite{DL,KLS,tach}. From this time on, the 
universe can be described by the usual big bang theory.

The main difference between inflationary theory and the old cosmology
becomes clear when one calculates the size of a typical inflationary
domain at the end of inflation. Investigation of this question    shows
that even if  the initial size of   inflationary universe  was as small
as the Plank size $l_P \sim 10^{-33}$ cm, after $10^{-35}$ seconds of
inflation   the universe acquires a huge size of   $l \sim 10^{10^{12}}$
cm! This number is model-dependent, but in all realistic models the  size of
the universe after inflation appears to be many orders of magnitude
greater than the size of the part of the universe which we can see now,
$l \sim 10^{28}$ cm. This immediately solves most of the problems of the
old cosmological theory \cite{Chaot,book}.

Our universe is almost exactly homogeneous on  large scale because all
inhomogeneities were exponentially stretched during inflation.  The
density of  primordial monopoles  and other undesirable ``defects''
becomes exponentially diluted by inflation.   The universe   becomes
enormously large. Even if it was a closed universe of a size
 $\sim 10^{-33}$ cm, after inflation the distance between its ``South'' and
``North'' poles becomes many orders of magnitude greater than $10^{28}$
cm. We see only a tiny part of the huge cosmic balloon. That is why
nobody  has ever seen how parallel lines cross. That is why the universe
looks so flat.

If our universe initially consisted of many domains
with chaotically distributed scalar field  $\phi$ (or if one considers
different universes with different values of the field), then  domains in
which the scalar field was too small never inflated. The main
contribution to the total volume of the universe will be given by those
domains which originally contained large scalar field $\phi$. Inflation
of such domains creates huge homogeneous islands out of initial chaos.
Each  homogeneous domain in this scenario is much greater than the size
of the observable part of the universe.

 \begin{figure}[h!]
\centering\leavevmode\epsfysize=5cm \epsfbox{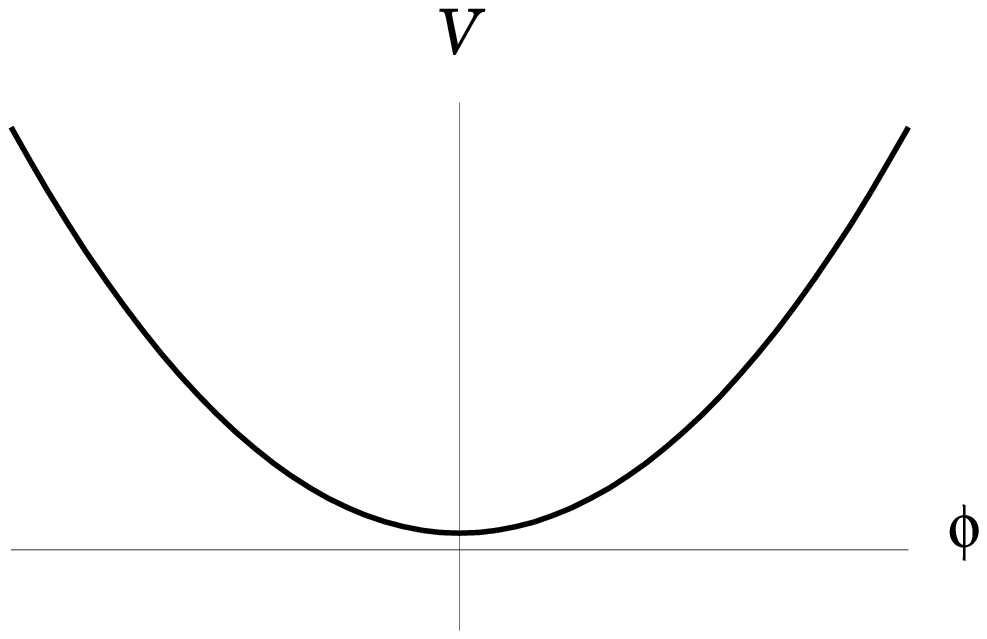} 
\caption[fig1]
{The harmonic oscillator scalar field potential $V(\phi) = V_0 + m^2\phi^2/2$ in a simplest version of chaotic inflation.}
\label{Cyclic7}
\end{figure}

\begin{figure}[b]
\centering\leavevmode\epsfysize=4cm  \epsfbox{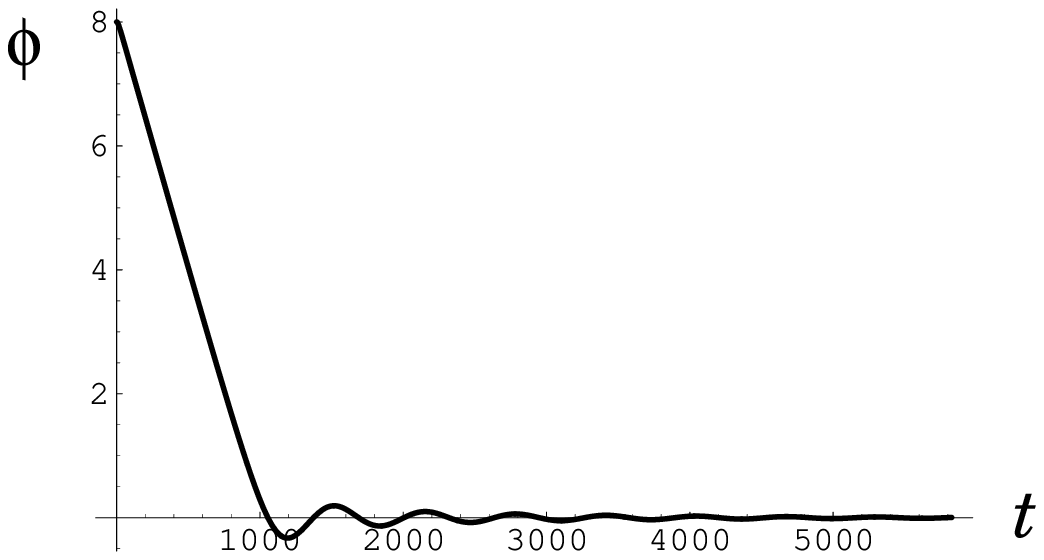}\\ \hskip -1 cm \centering\leavevmode\epsfysize=4.3cm ~~~\epsfbox{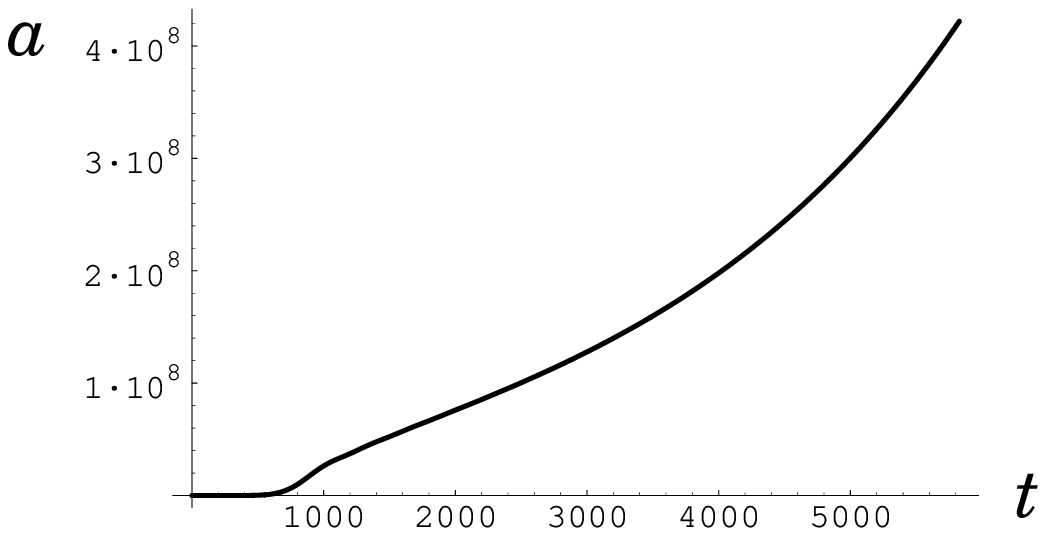}

\

\caption[fig1]{\label{qfigplus} Evolution of
the scalar field and the scale factor in the model $V(\phi) =
{m^2\over 2} \phi^2 + V_0 $ with $V_0 > 0$. In the beginning we have a
stage of inflation with the field $\phi$ linearly decreasing at $\phi
> 1$.  Then
the field enters a stage of oscillations with a gradually decreasing
amplitude of the field. When the energy of the
oscillations becomes smaller than $V_0$, the universe enters a second
stage of inflation, which corresponds to the present stage of acceleration of the universe.}
\end{figure}

Now let us make another simple step and add a small constant term $V_0>0$ to the potential $m^2\phi^2/2$, see Fig. \ref{Cyclic7}.
If one does not take gravity into account, this term does not change equations for the scalar field, i.e. we still consider a simplest harmonic oscillator theory. However, the term $V_0$ acts as a cosmological constant that does not vanish even at $\phi = 0$. As a result, the simplest theory with $V(\phi) = V_0 + m^2\phi^2/2$ will describe {\it two stages of inflation}. The first stage occurs in the early universe, when the scalar field was large. The second stage occurs right now; it corresponds to the recently discovered accelerated expansion of the universe, see Fig. \ref{qfigplus}.

The first models of chaotic inflation were based on the theories with
polynomial potentials, such as $V(\phi) = \pm {m^2\over 2} \phi^2
+{\lambda\over 4} \phi^4$. But the main idea of this scenario is quite
generic. One should consider any particular potential $V(\phi)$,
polynomial or not, with or without spontaneous symmetry breaking, and
study all possible initial conditions without assuming that the universe
was in a state of thermal equilibrium, and that the field $\phi$ was in
the minimum of its effective potential from the very beginning
\cite{Chaot}. This scenario strongly deviated from the standard lore of
the hot big bang theory and was psychologically difficult to accept.
Therefore during the first few years after invention of chaotic inflation
many authors claimed that the idea of chaotic initial conditions is
unnatural, and made attempts to realize the new inflation scenario based
on the theory of high-temperature phase transitions, despite numerous
problems associated with it.  Some authors even introduced so-called `thermal constraints' which were necessary to ensure that the minimum of the effective potential at large $T$ should be at $\phi=0$ \cite{OvrStein}, even though the scalar field in the models  they considered was not in a state of thermal equilibrium with other particles. Gradually, however,  it became clear that the
idea of chaotic initial conditions is most general, and   it is much
easier to construct a consistent cosmological theory without making
unnecessary assumptions about thermal equilibrium and high temperature
phase transitions in the early universe.

\section{Initial conditions for inflation \label{IniCond}}
Many other versions of inflationary cosmology have been proposed since
1983. Most of them are based not  on the  theory of high-temperature
phase transitions, as in old and new inflation, but on the idea of
chaotic initial conditions, which is the definitive feature of the
chaotic inflation scenario. The issue of initial conditions in inflationary cosmology was discussed by various methods including Euclidean quantum gravity, stochastic approach to inflation, etc, see e.g. \cite{book}. Here I would like to take a simple intuitive approach.

Consider  a closed universe of initial  size $l \sim 1$
(in Planck units), which emerges  from the
space-time foam, or from singularity, or from `nothing'  in a state
with  the Planck density $\rho \sim 1$. Only starting from this moment, i.e. at $\rho
\lesssim 1$, can we describe this domain as  a {\it classical} universe.  Thus,
at this initial moment the sum of the kinetic energy density, gradient energy
density, and the potential energy density  is of the order unity:\, ${1\over
2} \dot\phi^2 + {1\over 2} (\partial_i\phi)^2 +V(\phi) \sim 1$.

We wish to emphasize, that there are no {\it a priori} constraints on
the initial value of the scalar field in this domain, except for the
constraint ${1\over 2} \dot\phi^2 + {1\over 2} (\partial_i\phi)^2 +V(\phi) \sim
1$.  Let us consider for a moment a theory with $V(\phi) = const$. This theory
is invariant under the shift  $\phi\to \phi + a$. Therefore, in such a
theory {\it
all} initial values of the homogeneous component of the scalar field
$\phi$ are equally probable. 
The only constraint on the average amplitude of the field appears if
the effective potential is not constant, but grows and becomes greater
than the Planck density at $\phi > \phi_p$, where  $V(\phi_p) = 1$. This
constraint implies that $\phi \lesssim \phi_p$, but it does not give any reason to
expect that $\phi \ll \phi_p$. This suggests that the  typical initial value
$\phi_0$ of the field $\phi$ in such a  theory is  $\phi_0 \sim  \phi_p$. 
Thus, we expect that typical initial conditions correspond to
${1\over 2}
\dot\phi^2 \sim {1\over 2} (\partial_i\phi)^2\sim V(\phi) = O(1)$.
Note that if
by any chance ${1\over 2} \dot\phi^2 + {1\over 2} (\partial_i\phi)^2
\lesssim
V(\phi)$ in the domain under consideration, then inflation begins,
and  within
the Planck time the terms  ${1\over 2} \dot\phi^2$ and ${1\over 2}
(\partial_i\phi)^2$ become much smaller than $V(\phi)$, which ensures
continuation of inflation.  It seems therefore that chaotic inflation
occurs
under rather natural initial conditions, if it can begin at $V(\phi)
\sim 1$
\cite{Chaot,book}.

The assumption that inflation may begin at a very large $\phi$ has
important
implications. For example, in the theory $m^2\phi^2/2$ one has
$\phi_0 \sim  \phi_p \sim m^{-1}$.
Then, according to (\ref{E05}), the total size of a closed universe of a typical
initial size
$O(1)$  after
inflation becomes equal to
\begin{equation}\label{E09}
l \sim  e^{\phi^2_p/4}
\sim  e^{1/4m^2}  \ .
\end{equation}
For $m\sim 3\times 10^{-6}$ (which is necessary to produce
density perturbations $\delta{\rho}/\rho \sim 10^{-5}$, see
below)
\begin{equation}\label{E10}
l \sim   10^{10^{10}} cm \ .
\end{equation}
According to this estimate, the smallest possible domain of the
universe of
initial size  $O(M_p^{-1}) \sim 10^{-33}$ cm 
 after inflation becomes
much larger
than the size of the observable part of the universe $\sim 10^{28}
cm$\@. This is
the reason why our part of the universe looks flat, homogeneous and
isotropic.
The same mechanism solves also the horizon problem. Indeed, any
domain of
the Planck size, which becomes causally connected within the Planck
time, gives
rise to the part of the universe which is much larger than the part
which we
can see now.

In fact, as we will see in Section VI, once inflation begins in an interval $m^{-1/2} < \phi_0 < m^{-1}$ in the theory $m^2\phi^2/2$, the universe enters eternal process of self-reproduction \cite{Eternal}. Thus, if inflation begins in a single domain of a smallest possible size  $l = O(1)$, it makes the universe locally homogeneous and produces infinitely many inflationary domains of exponentially large size.

\section{Hybrid inflation}
In the previous section we considered the simplest chaotic inflation theory based on the theory of a single scalar field $\phi$. The models of chaotic inflation based on the theory of two scalar fields may have some qualitatively new features. One of the most interesting models of this kind is the hybrid inflation  scenario
\cite{Hybrid}.  The simplest version of this scenario is based on chaotic
inflation in the theory of two scalar fields with the effective potential
\begin{equation}\label{hybrid}
V(\sigma,\phi) =  {1\over 4\lambda}(M^2-\lambda\sigma^2)^2
+ {m^2\over
2}\phi^2 + {g^2\over 2}\phi^2\sigma^2\ .
\end{equation}
 The effective
mass squared of the field $\sigma$ is equal to  $-M^2 + g^2\phi^2$.
Therefore
for $\phi > \phi_c = M/g$ the only   minimum of the effective
potential
$V(\sigma,\phi)$ is at $\sigma = 0$. The curvature of the effective
potential in the $\sigma$-direction is much greater than in the
$\phi$-direction. Thus  at the first stages of
expansion of the universe the field $\sigma$ rolled down to $\sigma =
0$, whereas the field $\phi$ could remain large for a much longer
time.

At the moment when the inflaton field
$\phi$ becomes smaller than  $\phi_c = M/g$, the phase transition
with the
symmetry breaking occurs. If $m^2 \phi_c^2 = m^2M^2/g^2 \ll
M^4/\lambda$, the Hubble constant at the time of the phase transition
is given by $H^2 = {M^4 \over 12 \lambda}$ (in units $M_ = 1$).
If one assumes that $M^2 \gg {\lambda m^2\over g^2}$ and that
$m^2 \ll H^2$, then the universe at
$\phi > \phi_c$ undergoes a stage of inflation, which abruptly ends at $\phi =
\phi_c$.

One of the advantages of this scenario is the possibility to obtain small
density perturbations even if coupling constants are large, $\lambda, g  =
O(1)$. This scenario works   if the effective potential has a relatively flat
$\phi$-direction. But flat directions often appear in supersymmetric theories.
This makes hybrid inflation an attractive playground for those who wants to
achieve inflation in supergravity.

Another advantage of this scenario is a possibility to have inflation at $\phi \sim M/g 
\ll 1$. (This happens because the slow rolling of the field $\phi$ in this scenario is supported not by the energy of the field $\phi$ as in the scenario described in the previous section, but by the energy of the field $\sigma$.) This   helps to avoid problems which may appear if the effective
potential in supergravity and   string theory blows up at $\phi > 1$.
Several different models of hybrid inflation in supergravity have been proposed
during the last few  years (F-term  inflation \cite{F}, D-term inflation
\cite{D},    etc.) A detailed discussion of various versions of hybrid
inflation in supersymmetric theories can be found in \cite{LythRiotto}. Recent developments in this direction have been reported by Kallosh at this conference \cite{renata}.

\section{Quantum fluctuations and density perturbations
\label{Perturb}}

The vacuum structure in the  exponentially expanding universe  is much more complicated than in ordinary Minkowski space.
 The wavelengths of all vacuum
fluctuations of the scalar field $\phi$ grow exponentially
during inflation. When the wavelength of any
particular fluctuation becomes greater than $H^{-1}$, this
fluctuation stops oscillating, and its amplitude freezes at
some nonzero value $\delta\phi (x)$ because of the large
friction term $3H\dot{\phi}$ in the equation of motion of the field
$\phi$\@. The amplitude of this fluctuation then remains
almost unchanged for a very long time, whereas its
wavelength grows exponentially. Therefore, the appearance of
such a frozen fluctuation is equivalent to the appearance of
a classical field $\delta\phi (x)$ that does not vanish
after averaging over macroscopic intervals of space and
time.

Because the vacuum contains fluctuations of all
wavelengths, inflation leads to the creation of more and
more new perturbations of the classical field with
wavelengths greater than $H^{-1}$\@. The average amplitude of
such perturbations generated during a typical time interval $H^{-1}$ is given
by \cite{Vilenkin:wt,Linde:uu}
\begin{equation}\label{E23}
|\delta\phi(x)| \approx \frac{H}{2\pi}\ .
\end{equation}
 
These fluctuations lead to density perturbations that later produce galaxies. The theory of this effect is very complicated \cite{Mukh,Hawk}, and it was fully understood only in the second part of the 80's \cite{Mukh2}. Here we will only give a rough and oversimplified idea of this effect. 

Fluctuations of the field $\phi$ lead to a local delay of the time of the end of inflation, $\delta t = {\delta\phi\over \dot\phi} \sim {H\over 2\pi \dot \phi}$. Once the usual post-inflationary stage begins, the density of the universe starts to decrease as $\rho = 3 H^2$, where $H \sim t^{-1}$. Therefore a local delay of expansion leads to a local density increase $\delta_H$ such that $\delta_H \sim \delta\rho/\rho \sim  {\delta t/t}$. Combining these estimates together yields the famous result \cite{Mukh,Hawk,Mukh2}
\begin{equation}\label{E24}
\delta_H \sim \frac{\delta\rho}{\rho} \sim {H^2\over 2\pi\dot\phi} \ .
\end{equation}
This  derivation is oversimplified; it does not tell, in particular, whether $H$ should be calculated during inflation or after it. This issue was not very important for new inflation where $H$ was nearly constant, but it is of crucial importance for chaotic inflation.

The result of a more detailed investigation \cite{Mukh2} shows that $H$ and $\dot\phi$ should be calculated during inflation, at different times for perturbations with different momenta $k$. For each of these perturbations the value of $H$ should be taken at the time when the wavelength of the perturbation  becomes of the order of $H^{-1}$. However, the field $\phi$ during inflation changes very slowly, so the quantity ${H^2\over 2\pi\dot\phi}$ remains almost constant over exponentially large range of wavelengths. This means that the spectrum of perturbations of metric is flat. 

A detailed calculation in our simplest chaotic inflation model the amplitude of perturbations gives \cite{LythRiotto}
\begin{equation}\label{E26}
\delta_H \sim   {m \phi^2\over 5\pi \sqrt 6} \ .
\end{equation}
The perturbations on scale of the horizon were produced at $\phi_H\sim 15$ \cite{book}. This, together with COBE normalization $\delta_H \sim 2 \times 10^{-5}$  gives $m \sim 3\times 10^{-6}$, in Planck units, which is approximately equivalent to
$7 \times 10^{12}$ GeV. Exact numbers depend on $\phi_H$, which in its turn depends slightly on the subsequent thermal history of the universe.

The magnitude of density perturbations $\frac{\delta \rho}{\rho}$ in our model depends on the scale $l$ only logarithmically. Flatness of the spectrum of $\frac{\delta \rho}{\rho}$  together with flatness of the universe ($\Omega = 1$)  constitute the two most robust predictions of inflationary cosmology. It is possible to construct models where $\frac{\delta \rho}{\rho}$ changes in a very peculiar way, and it is also possible to construct theories where $\Omega \not = 1$, but it is difficult to do so.

\section{From the Big Bang theory to the theory of eternal inflation}

The next step in the development of inflationary theory that I would like to discuss here is the discovery of the process of self-reproduction of inflationary universe. This process was known to exist in old inflationary theory \cite{Guth} and in the new one \cite{Vilenkin:xq}, but it is especially surprising and leads to most profound consequences in the context of the chaotic inflation scenario \cite{Eternal,LLM}. It appears that   large scalar field during inflation produces large quantum fluctuations which may locally increase the value of the scalar field in some parts of the universe. These regions expand at a greater rate than their parent domains, and quantum fluctuations inside them lead to production of new inflationary domains that expand even faster. This surprising behavior leads to an eternal process of self-reproduction of the universe.

To understand the mechanism of self-reproduction, remember that
the processes separated by distances $l$ greater than $H^{-1}$
proceed independently of one another. Indeed,
during exponential expansion the distance between any two objects
separated by more than $H^{-1}$ grows with a
speed exceeding the speed of light.
As a result, an observer in the inflationary
universe can see only the processes occurring inside the
horizon of the radius  $H^{-1}$. An important consequence of this general result is that the process of inflation in any domain of radius
$H^{-1}$ occurs independently of any events outside it.
In this sense any inflationary domain of
initial radius exceeding $H^{-1}$ can be considered as a
separate mini-universe.  

To investigate the behavior of such a mini-universe, with an
account taken of quantum fluctuations, let us consider an
inflationary domain of initial radius $H^{-1}$
containing sufficiently homogeneous field with initial
value $\phi \gg 1$.
Equation (\ref{E04}) implies that during a typical time interval $\Delta t=H^{-1}$ the field
inside this domain will be reduced by
$\Delta\phi = \frac{V'}{V}= 2/\phi$. By comparison of this expression with
$|\delta\phi(x)| \approx \frac{H}{2\pi} = \sqrt{V(\phi) \over 2\pi\sqrt 3 } \sim {m\phi\over 2\pi\sqrt 6}$  one can easily see
that if $\phi$ is much greater than $\phi^* \sim {5 \over \sqrt{ m}} $,
 then the decrease of the field $\phi$
due to its classical motion is much smaller than the average
amplitude of the quantum fluctuations $\delta\phi$ generated
during the same time. Because the
typical wavelength of the fluctuations $\delta\phi (x)$
generated during the time is $H^{-1}$, the whole domain
after $\Delta t = H^{-1}$ effectively becomes divided
into $e^3 \sim 20$ separate domains (mini-universes) of radius
$H^{-1}$, each containing almost homogeneous field $\phi -
\Delta\phi+\delta\phi$.   In almost a half of these domains   the field $\phi$ grows by $|\delta\phi(x)|-\Delta\phi
\approx
|\delta\phi (x)| = H/2\pi$, rather than decreases. This means that the total volume of the universe containing {\it growing} field $\phi$ increases 10 times. During the next
time interval
$\Delta t = H^{-1}$ the situation repeats. Thus, after the two time  intervals $H^{-1}$
the total volume of the universe containing the growing scalar field increases 100 times, etc.   The universe enters eternal process of self-reproduction. 
Note that this process begins at $V(\phi) \sim m  \ll 1$, i.e. at a density that is still  much smaller than the Planck density.
In a more general case, the criterion for self-reproduction is $12\pi^2 V'^2 \ll V^3$ \cite{Eternal,LLM}.

Until now we have considered the simplest 
inflationary model with only one scalar field, which had only one minimum of its potential energy. Meanwhile, realistic 
models of elementary particles describe many kinds of scalar fields.  The potential energy of these scalar fields may have several different minima. This means that the same theory may have different ``vacuum states," corresponding to different types of symmetry breaking between fundamental interactions, and, consequently, to different laws of low-energy physics.

    As a result of quantum jumps of the scalar fields during inflation, the universe may become divided into infinitely many exponentially large 
domains that have different laws of low-energy physics. Note that 
this division occurs even if the whole universe originally began in 
the same state, corresponding to one particular minimum of 
potential energy.

To illustrate this scenario, we present here the results of computer simulations of evolution of a system of two scalar fields during inflation. The field $\phi$ is the inflaton field driving inflation; it is shown by the height of the distribution of the field $\phi(x,y)$ in a two-dimensional slice of the universe. The second field, $\Phi$, determines the type of spontaneous symmetry breaking which may occur in the theory. We paint the surface black if this field is in a state corresponding to one of the two minima of its effective potential;  we paint it white if it is in the second minimum corresponding to a different type of symmetry breaking, and therefore to a different set of laws of low-energy physics.

\begin{figure}

\centering\leavevmode\epsfysize=7.5cm \epsfbox{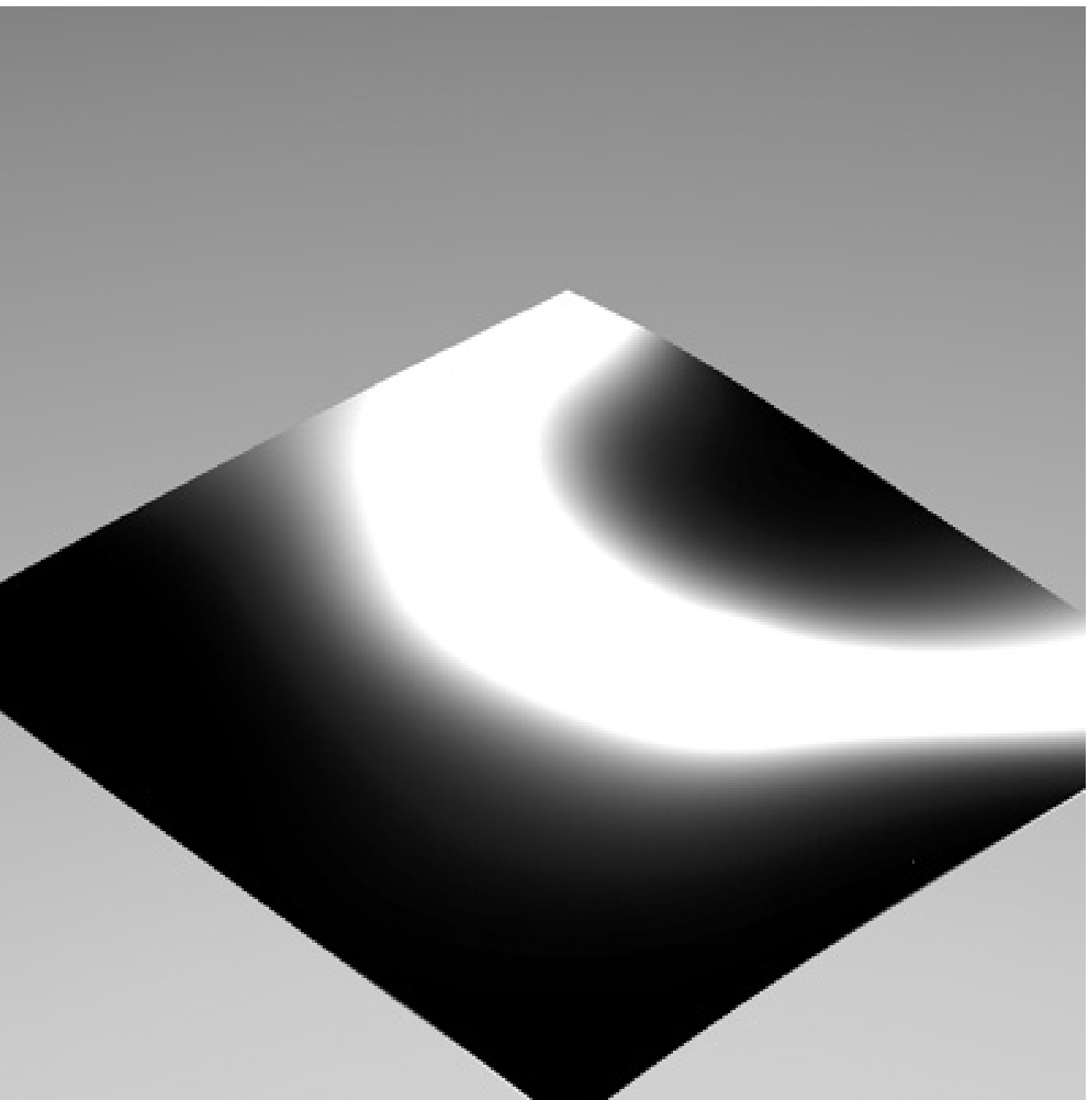}\\ 
 
 \vskip 0.25cm

  \centering\leavevmode\epsfysize=7.5cm\epsfbox{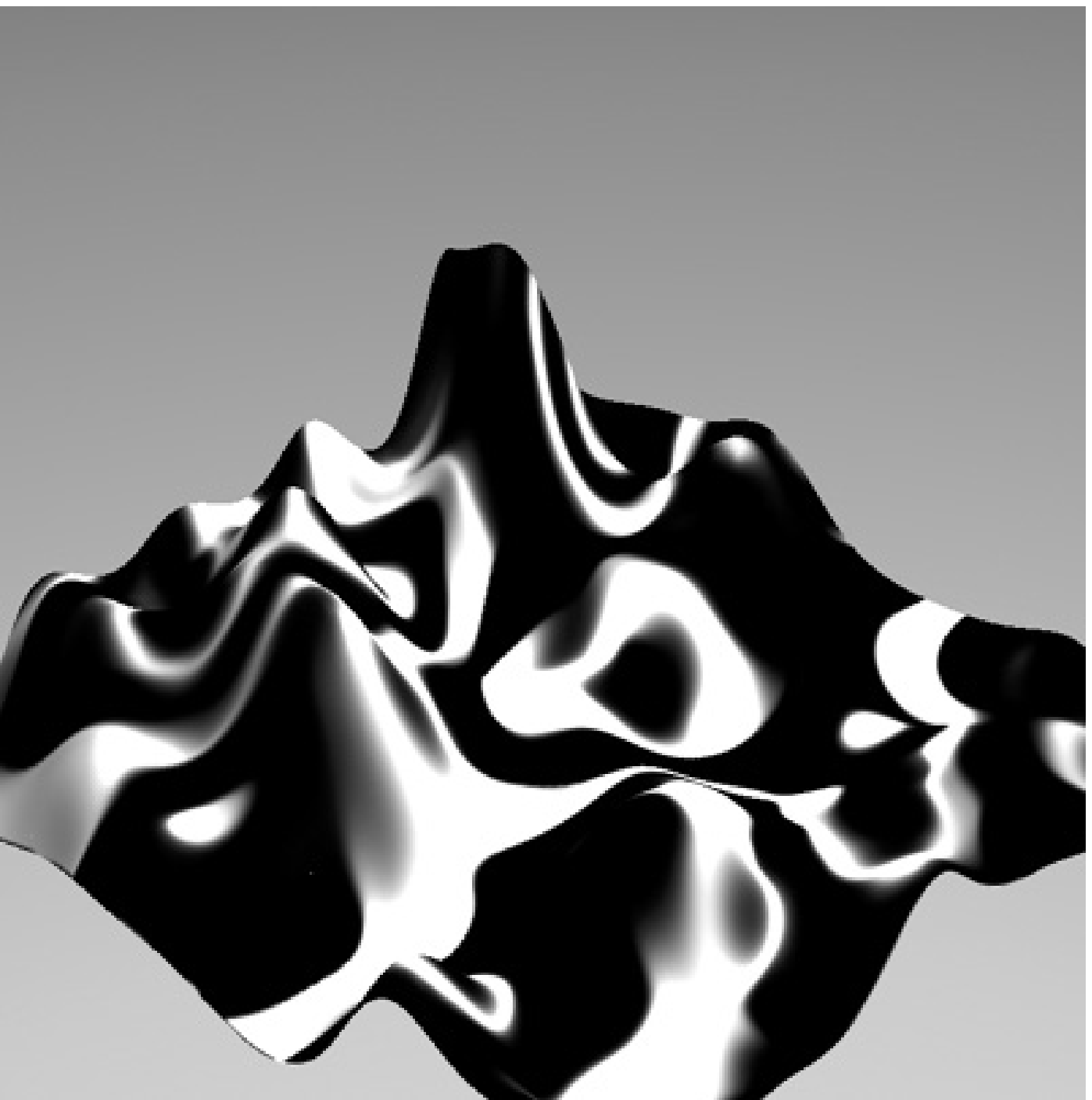}\\
  
 \vskip 0.25cm

\centering\leavevmode\epsfysize=7.5cm \epsfbox{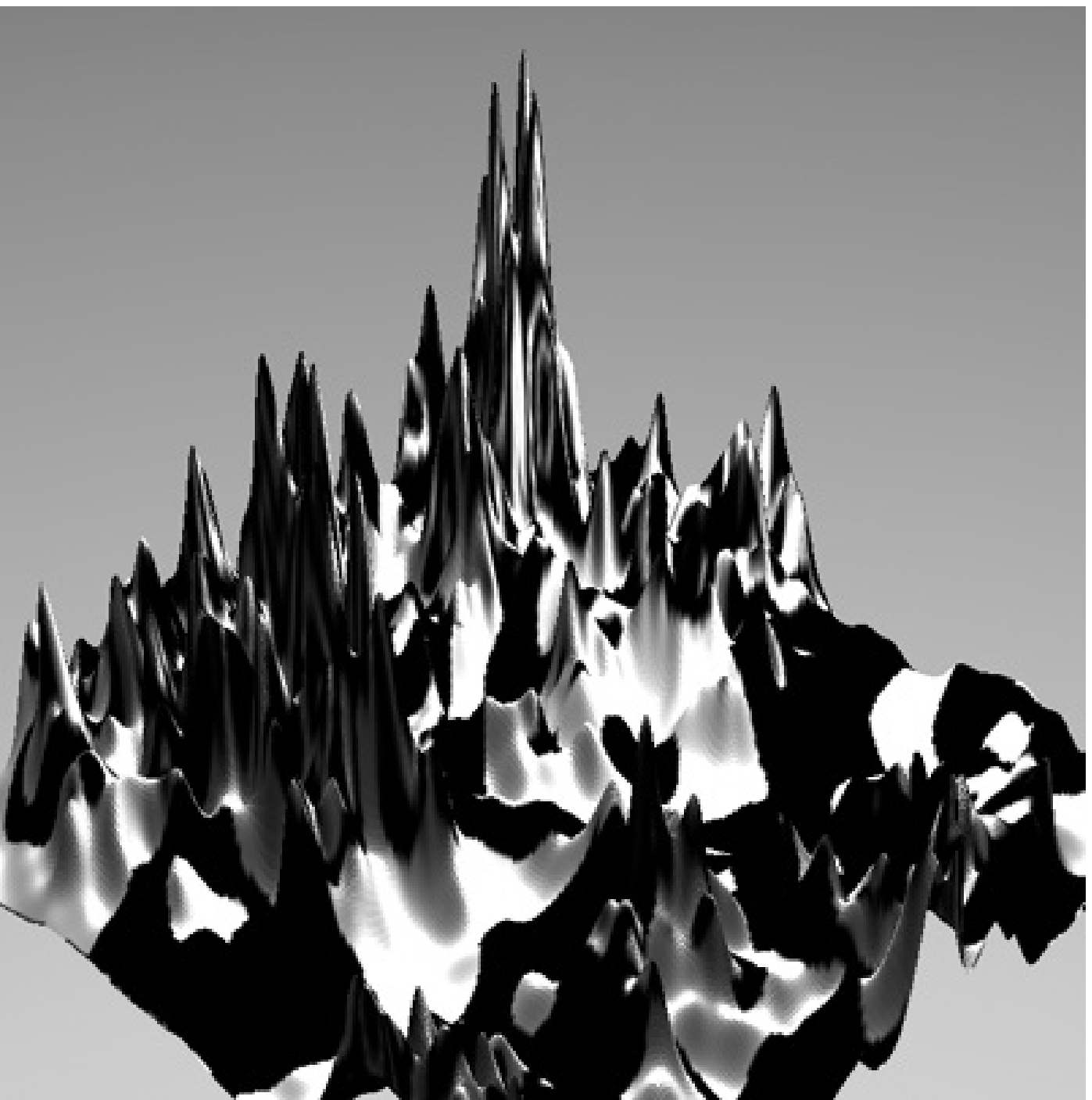}\\

\hskip -0.1cm  \centering\leavevmode\epsfysize=8.3cm\epsfbox{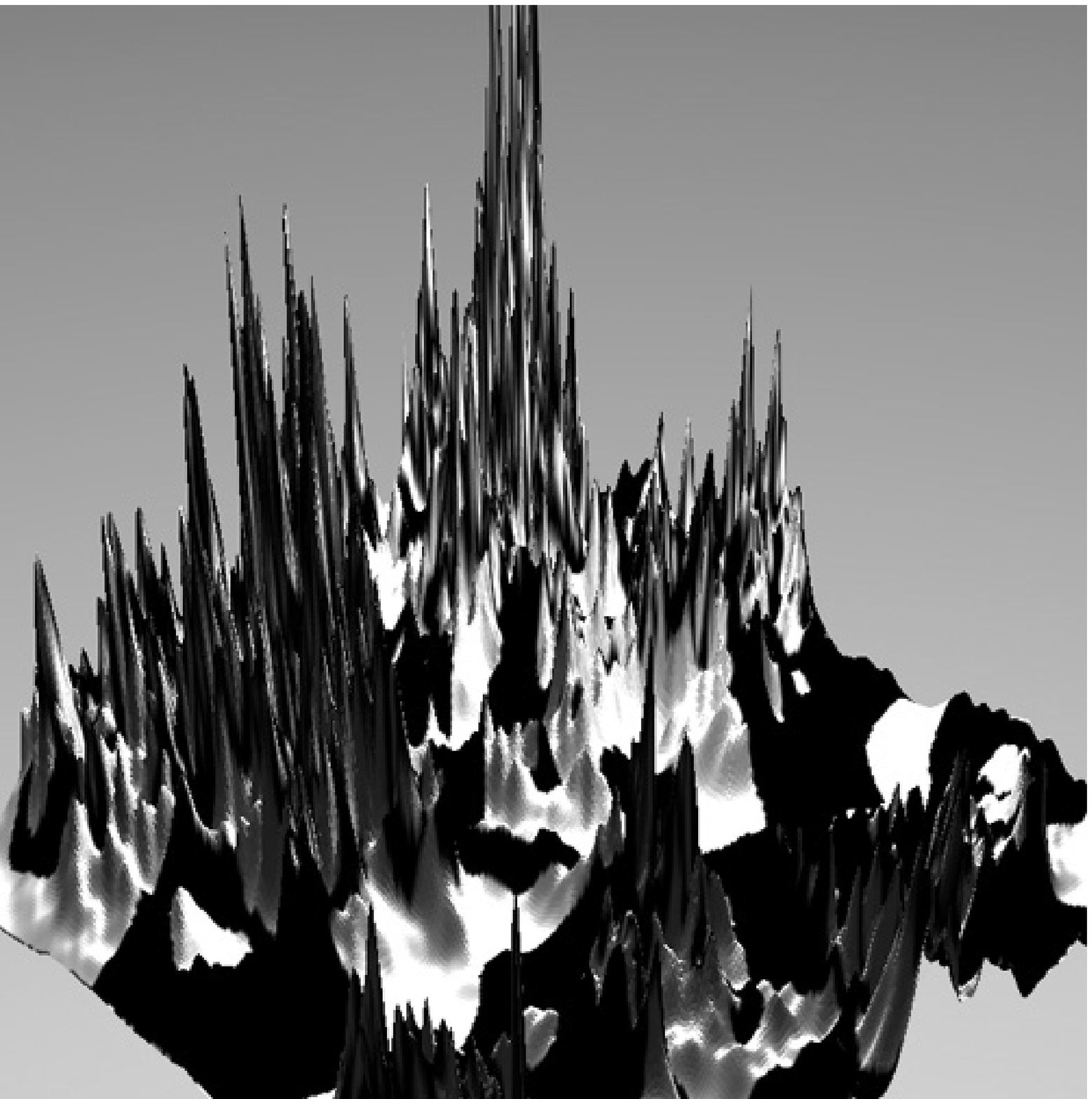}

\

\caption{Evolution of scalar fields  $\phi$ and $\Phi$ during the process of self-reproduction of the universe.   The height of the distribution shows the value of the field $\phi$ which drives inflation. The surface is painted black in those parts of the universe where the scalar field $\Phi$ is in the first minimum of its effective potential, and  white where it is in the second minimum. Laws of low-energy physics are different in the regions of different color. The peaks of the ``mountains'' correspond  to places where quantum fluctuations bring the scalar fields back to the Planck density. Each of such places in a certain sense can be considered as a beginning of a new Big Bang. }
\label{fig:Fig0}
\end{figure}

In the beginning of the process the whole inflationary domain is black, and the distribution of both fields is very homogeneous. Then the domain became exponentially large (but it has the same size in comoving coordinates, as shown in Fig. \ref{fig:Fig0}).  Each peak of the mountains corresponds to nearly Planckian density and can be interpreted as a beginning of a new ``Big Bang.'' The laws of physics are rapidly changing there, but they become fixed in the parts of the universe where the field $\phi$ becomes small. These parts correspond to valleys in Fig. \ref{fig:Fig0}. Thus quantum fluctuations of the scalar fields divide the universe into exponentially large domains with different laws of low-energy physics, and with different values of energy density. 

 Note, that this process occurs only if the Hubble constant during inflation is much greater than the masses of the field $\Phi$ in the minima of the effective potential. In the new inflation scenario the Hubble constant $H$ was about three orders of magnitude smaller than $M_{\rm GUT}$ and six orders of magnitude smaller than the Planck mass. The transitions of the type discussed above would be impossible of at least very improbable.
This was one of the reasons why the realization that new inflation is eternal did not attract much interest and for a long time remained essentially forgotten by everyone including those who have found this effect \cite{Vilenkin:xq}.

The situation changed completely when it was found that eternal inflation occurs in the chaotic inflation scenario \cite{Eternal}. Indeed, in this case eternal process of self-reproduction of the universe  is possible even at $H \sim M_p$. This allows the transitions between all vacua even if the masses of the corresponding scalar fields approach the Planck mass. In such a case the universe can probe all possible vacuum states of the theory. This for the first time provided physical justification of the anthropic principle.

Indeed, if this scenario is correct, then physics alone cannot provide a 
complete explanation for all properties of our part of the universe.   
The same physical theory may yield large parts of the universe that 
have diverse properties.  According to this scenario, we find 
ourselves inside a four-dimensional domain with our kind of 
physical laws not because domains with different dimensionality 
and with alternate properties are impossible or improbable, but 
simply because our kind of life cannot exist in other domains. 
 
\section{Inflation and observations}

Inflation is not just an interesting theory that can resolve many difficult problems of the standard Big Bang cosmology. This  theory made several   predictions which can be tested by cosmological observations. Here are the most important predictions: 

1) The universe must be flat. In most models $\Omega_{total} = 1 \pm 10^{-4}$.

2) Perturbations of metric produced during inflation are adiabatic.  

3) Inflationary perturbations have flat spectrum.  In most inflationary models the spectral index  $n = 1 \pm 0.2$ ($n=1$ means totally flat.)

4) These perturbations are gaussian.  

5) Perturbations of metric could be scalar, vector or tensor. Inflation mostly produces scalar perturbations, but it also produces tensor perturbations with nearly flat spectrum, and it does {\it not} produce vector perturbations. There are certain relations between the properties of  scalar and tensor perturbations produced by inflation.

6) Inflationary perturbations produce specific peaks in the spectrum of CMB radiation.

It is possible to violate each of these predictions if one makes this theory sufficiently complicated. For example, it is possible to produce vector perturbations of metric in the models where  cosmic strings are produced at the end of inflation, which is the case in some versions of hybrid inflation. It is possible to have open or closed inflationary universe, it is possible to have models with nongaussian isocurvature fluctuations with a non-flat spectrum. However, it is extremely difficult to do so, and most of the inflationary models satisfy the simple rules given above.  

It is not easy to test all of these predictions. The major breakthrough in this direction was achieved due to the recent measurements of the CMB anisotropy. These measurements revealed the existence of two (or perhaps even three) peaks in the CMB spectrum. Position of these peaks is consistent with predictions of the simplest inflationary models with adiabatic gaussian perturbations, with $\Omega = 1 \pm 0.03$, and $n = 1.03 \pm 0.06$~\cite{Bond}.

Inflationary
scenario is very versatile, and now, after 20 years of persistent
attempts of many physicists to propose an alternative to inflation, we
still do not know any other  way to construct a consistent cosmological
theory.  But may be we  did not try hard enough? 

Since most of inflationary models are based on 4D cosmology, it would be natural to venture into the study of higher-dimensional cosmological models. In what follows we will discuss one of the recent attempts to formulate an alternative cosmological scenario.

\section{Alternatives to inflation?}
There were many attempts to suggest an alternative to inflation. However, in order to compete with inflation a new theory should offer an alternative solution to many difficult cosmological problems. Let us look at these problems again before starting a discussion.

1) Homogeneity problem. Before even starting investigation of density perturbations and structure formation, one should explain why the universe is nearly homogeneous on the horizon scale.

2) Isotropy problem. We need to understand why all directions in the universe are similar to each other, why there is no overall rotation of the universe. etc.

3) Horizon problem. This one is closely related to the homogeneity problem. If different parts of the universe have not been in a causal contact when the universe was born, why do they look so similar?

4) Flatness problem. Why $\Omega = 1 \pm 0.03$? Why parallel lines do not intersect?

5) Total entropy problem. The total entropy of the observable part of the universe is greater than $10^{87}$. Where did this huge number come from? Note that the lifetime of a closed universe filled with hot gas with total entropy $S$  is $S^{2/3}\times 10^{-43}$ seconds \cite{book}. Thus $S$ must be huge. Why?

6) Total mass problem. The total mass of the observable part of the universe has mass $\sim 10^{60} M_p$.  Note also that the lifetime of a closed universe filled with nonrelativistic particles of total mass $M$ is ${M\over M_P} \times 10^{-43}$ seconds. Thus $M$ must be huge. But why?

7) Structure formation problem. If we manage to explain the homogeneity of the universe, how can we explain the origin of inhomogeneities required for the large scale structure formation?

8) Monopole problem, gravitino problem, etc.

This list is very long. That is why it was not easy to propose any alternative to inflation even before we learned that $\Omega \approx 1$, $n\approx 1$, and that the perturbations responsible for galaxy formation are mostly adiabatic, in agreement with predictions of the simplest inflationary models.

Despite this difficulty (or maybe because of it) there was always a tendency to announce that we have eventually found a good alternative to inflation. This was
the ideology of the models of structure formation due to topological
defects or textures. Of course, even 10 years ago everybody knew that these theories at best could solve only one problem (structure formation) out of the 8 problems mentioned above. The true question was whether inflation  with cosmic strings/textures is any better than inflation  without cosmic strings/textures. However, such a formulation would not make the headlines.  Therefore the models of structure formation due to topological defects or textures sometimes were advertised in public press as the models that ``match the explanatory triumphs of inflation while rectifying its major
failings'' \cite{SperTur}.

Recently the theory of topological defects and textures as a source of the large scale structure was essentially ruled out by observational data, but the tradition of advertisement of various `successful' alternatives to inflation is still flourishing. Recent example is given by the ekpyrotic/cyclic scenario \cite{KOST,cyclic}. The  50-pages long paper on ekpyrotic scenario \cite{KOST} appeared in hep-th in April 2001, and ten days later, before any experts could make their judgement, it was already enthusiastically discussed on BBC and CNN as a viable alternative to inflation. The reasons for the enthusiasm can be easily understood. We were told that finally we have a cosmological theory that is based on string theory and that is capable of solving all major cosmological problems without any use of inflation, which was called   `superluminal expansion' \cite{KOST}.

However, a first look at this scenario  revealed several minor problems and inconsistencies, then we have found much more serious problems, and eventually it became apparent that the original version of the ekpyrotic scenario \cite{KOST} did not live to its promise \cite{KKL,KKLT}. In particular, instead of expanding, the ekpyrotic universe collapses to a singularity \cite{KKLT,Khoury:2001bz}. The theory of density perturbations in this scenario is very controversial \cite{Khoury:2001zk,Lyth:2001pf,Brandenberger:2001bs,Hwang:2002ks,Martin:2001ue,Durrer:2002jn}. More importantly, this scenario offers no solution to major cosmological problems such as the homogeneity, flatness and entropy problems, so it is not a viable alternative to inflation \cite{KKL}. Let me explain this part first, before we turn our attention to the cyclic scenario \cite{cyclic}.

\section{Ekpyrosis}
\subsection{Basic scenario}

According to the ekpyrotic scenario~\cite{KOST}, we live at one of the two `heavy' 4D branes in 5D universe described by the Ho\u{r}ava-Witten (HW) theory~\cite{HoravaWitten}. Our brane is called visible, and the second brane is called hidden.
There is also a  `light' bulk brane at a distance $Y$ from the visible brane  in the 5th direction.
 
The three brane configuration is assumed to be in a nearly stable BPS state. The
bulk brane has  potential energy   $V(Y)= -v e^{-m\alpha Y}$, where $m$ is some constant. It is assumed that at small $Y$ the potential suddenly vanishes.

The bulk brane moves towards our brane and collides with it. Due to the slight contraction of the scale factor of the universe, the bulk  brane carries some residual kinetic energy
immediately before the collision with the visible brane. After the
collision, this residual kinetic energy transforms into radiation which
will be deposited in the three dimensional space of the visible  brane.
The visible brane, now filled with hot radiation, somehow begins to expand
as a flat FRW universe. 
Quantum fluctuations of the position of the bulk brane generated during
its  motion from $Y= R$ to $Y=0$  will result in  density fluctuations
with a nearly flat spectrum.  The spectrum will have a blue tilt. It is argued that the problems of homogeneity, isotropy, flatness and horizon do not appear it this model because the universe, according to~\cite{KOST}, initially was in a nearly BPS state, which is homogeneous.

\subsection{Ekpyrotic scenario versus string theory}

It would be great to have a realistic brane cosmology based on string theory. However, ekpyrotic theory is just one of the many attempts to do so, and its relation to string theory is rather indirect.

One of the central assumptions of the ekpyrotic scenario is that we live on a negative tension brane. However, the standard HW phenomenology~\cite{HoravaWitten} is based  on the assumption that the tension of the visible brane is positive.  There were two main reasons for such an assumption. First of all, in practically all known  versions of the HW phenomenology,  with few exceptions,  a smaller group of symmetry (such as $E_6$) lives on the positive tension brane and provides the basis for GUTs, whereas the symmetry $E_8$ on the negative tension brane may remain unbroken. It is very difficult   to find  models where $E_6$ or $SU(5)$ live on the negative tension brane~\cite{Benakli:1999sy,Donagi:2001fs}.  

There is another reason why the tension of the visible brane is positive in the standard HW phenomenology~\cite{HoravaWitten}:  The square of the gauge coupling constant is inversely proportional to the Calabi-Yau volume~\cite{HoravaWitten}. On the negative tension brane this volume is greater than on the positive tension one, see e.g.~\cite{KOST}. In the standard HW phenomenology it is usually assumed that we live on the positive tension brane with small gauge coupling, ${g^2_{GUT}\over 4\pi} \sim 0.04$. On the hidden brane with negative tension the gauge coupling constant becomes large, ${g^2_{hidden}\over 4\pi} = O(1)$, which makes the gaugino condensation possible~\cite{HoravaWitten}. It is not  impossible to have a consistent phenomenology with the small gauge coupling on the hidden brane, but this is an unconventional and not well explored possibility~\cite{Benakli:1999sy}.  Therefore the original version of the ekpyrotic scenario was at odds with the standard HW phenomenology as defined in~\cite{HoravaWitten}.

Another set of problems is related to the potential $V(Y)$ playing crucial role in this scenario. This potential is supposed to appear as a result of nonperturbative effects. However, it is not clear whether the potential with required properties may actually emerge in the HW theory. Indeed, the potential $V(Y)$ must be very specific. It should vanish at $Y=0$, and it must be negative and behave as $- e^{-\alpha m Y}$ at large $Y$. One could expect terms like that, but in general one also obtains terms such as $\pm  e^{-\alpha m (R-Y)}$, where $R-Y$ is the distance between the bulk brane and the second `heavy' brane ~\cite{Moore:2000fs}. Such terms, as well as power-law corrections, must be forbidden if one wants to obtain density perturbations with flat spectrum~\cite{KKL}. The only example of a calculation of the potential of such type was given in~\cite{Moore:2000fs}. In this example the ``forbidden'' terms $\pm  e^{-\alpha m (R-Y)}$ do appear, and the sum of all terms is not negative, as assumed in \cite{KOST}, but strictly positive~\cite{Moore:2000fs}.

An additional important condition is that near the hidden brane the absolute value of the potential must be smaller than $e^{-120}$, because otherwise the density perturbations on the scale of the observable part of the universe will not be generated~\cite{KKL}. Also, if one adds a positive constant suppressed by the factor $\sim e^{-120}$ to $V(Y)$, inflation may begin. This is something the authors of the ekpyrotic scenario were trying to avoid.

But if the nonperturbative effects responsible for $V(Y)$ are so weak, how can they compete with the strong forces which are supposed to stabilize the positions of the visible brane and the hidden brane in Ho\u{r}ava-Witten theory? Until the brane stabilization mechanism is understood, it is very hard to trust any kind of ``derivation'' of the miniscule nonperturbative potential $V(Y)$ with extremely fine-tuned properties.

After we made these comments \cite{KKL}, the authors of the ekpyrotic scenario have changed it. In the cyclic scenario \cite{cyclic} the potential $V(Y)$ is supposed to approach a small {\it positive} value at large $Y$. This leads to inflation that is supposed to solve the homogeneity problem. Thus, cyclic scenario is no longer an alternative to inflation. Also, the branes in cyclic scenario are not stabilized. Therefore this scenario is no longer related to the standard string phenomenology.

\subsection{Singularity problem}

The discussion of brane collision in the ekpyrotic scenario was based on static 5D solution describing non-moving branes in the absence of the potential $V(Y)$ \cite{KOST}. However, the   action   as well as the solution  for the static 3-brane configuration given in \cite{KOST} was  not quite correct. The corrected  version of the action and  the  solution was given in ~\cite{KKLT}. 

More importantly, the  solution discussed there was given for branes that were not moving. It was assumed  in \cite{KOST} that in order to study 5D cosmological solutions it is sufficient to take the static metric used in \cite{KOST} and make its coefficients time-dependent.  However, we have found that the 5D cosmological solution given in \cite{KOST} was incorrect. The ansatz for the metric and the fields used in~\cite{KOST} does not solve the time-dependent 5D equations~\cite{KKLT}. Moreover, we have shown that if one uses effective 4D theory in order to describe the brane motion, as proposed in \cite{KOST}, the universe after the brane collision can only collapse ~\cite{KKLT}. Thus, instead of the Big Bang one gets a Big Crunch! This conclusion later was confirmed in  \cite{Khoury:2001bz}.

Since that time, the singularity problem became an unavoidable  part of the ekpyrotic/cyclic scenario. This problem is very  complicated. In this respect, various authors  have rather different opinions. Those who ever tried to solve this problem in general relativity are more than skeptical. The authors of the ekpyrotic scenario hope that experts in string theory will solve the problem of the cosmological singularity very soon. Meanwhile, the experts in string theory study toy models, such as the 2+1 dimensional  model  with a null singularity  rather than with a space-like cosmological singularity \cite{SeibergMoore}. Even in these models the situation is very complicated because of certain divergent scattering amplitudes. Therefore  at present string theorists do not want to make any speculations about the resolution of the singularity problem in realistic cosmological theories \cite{Nekrasov,Seiberg}.

\subsection{Density perturbations}
The problems discussed above are extremely complicated. But let us take a positive attitude and assume for a moment that all of these problems eventually will be solved. 

Now let us discuss the mechanism of generation of density perturbations in the ekpyrotic scenario. As was shown in~\cite{KKL}, this mechanism is based on the tachyonic instability with respect to generation of quantum fluctuations of the bulk brane position in the theory with the potential $V(Y) \sim -e^{-\alpha m Y}$. One may represent the position of the brane $Y(x)$ by a scalar field $\phi$, and find  that the long wavelength quantum fluctuations of this field grow exponentially because the effective mass squared of this field, proportional to $V''(Y)$, is negative.    A detailed theory of such instabilities recently was developed in the context of the theory of tachyonic preheating~\cite{tach}.

Inhomogeneities of the brane position lead to the $x$-dependent time delay of the  Big Crunch, i.e. of the moment when the `brane damage' occurs and matter is created. In inflationary theory, a position-dependent delay of the moment of reheating leads to density perturbations~\cite{Mukh,Hawk}. Simple estimates based on a similar idea lead to the conclusion~\cite{KKL} that in the scenario with $V(Y) \sim -e^{-\alpha m Y}$ one obtains a nearly flat spectrum of perturbations $\delta t$. If one naively multiplies $\delta t$ by $H$ after the brane collision, these perturbations translate into density perturbations $\delta\rho/\rho \sim H\delta t$. These perturbations have flat spectrum with a small red tilt \cite{KKL}.\footnote{If one takes non-exponential potentials decreasing at large $Y$, such as $-Y^n$ with $n< 40$, the tilt becomes unacceptably large \cite{KKL}. This brings back the unsolved problem of the origin of the purely exponential potential $V(Y)$ postulated in \cite{KOST}.}

However, this approach is oversimplified. Just as in the case of inflationary theory, one should specify when the Hubble constant is to be evaluated. In the ekpyrotic theory this question is crucial, because during the process of production of the perturbations $\delta t$ the Hubble constant is vanishingly small, and it was rapidly changing. Therefore if one multiplies $\delta t$ by $H$ at the time of the production of fluctuations $\delta\phi$, as we did for inflationary theory, one will get extremely small perturbations with a non-flat spectrum.

In order to obtain an unambiguous result, one should use the methods developed in \cite{Mukh2}. There were several attempts to do so. The authors of the ekpyrotic theory, as usual, are very optimistic and claim that they are obtaining perturbations with flat spectrum \cite{Khoury:2001zk}. Meanwhile everybody else \cite{Lyth:2001pf,Brandenberger:2001bs,Hwang:2002ks,Martin:2001ue}, with exception of Ref. \cite{Durrer:2002jn},  insist  that adiabatic perturbations with flat spectrum are not generated in this scenario, or, in the best case, we simply cannot tell anything until the singularity problem in this theory is resolved.

I share this point of view, and I would like to add to it something else. The theory of density perturbations in the ekpyrotic/cyclic scenario is immensely complicated. The authors of  \cite{Khoury:2001zk} have already made 3 revisions of their paper, and the final results continue slightly changing. Whereas originally the spectrum of perturbations in their scenario was supposed to be blue (decreasing at large scales) \cite{KOST}, now they say that it is red. In order to obtain perturbations of desirable magnitude, it is usually required that the absolute value of the effective potential $V(Y)$ should be greater than the Planck density, $|V(Y)| \gtrsim 1$ \cite{Khoury:2001zk,cyclic}, see below. But in this case the perturbation theory is expected to fail, quite independently of the singularity problem.

 I believe that a complete analysis of this problem could be possible only in the 5D theory rather than in the effective 4D theory. This brings additional complications described in \cite{KKLT,Rasanen:2001hf}.

\subsection{And the main problem is...} 

If adiabatic density perturbations with flat spectrum are not produced \cite{Lyth:2001pf,Brandenberger:2001bs,Hwang:2002ks,Martin:2001ue}, one may stop any further discussion of this scenario. However, let us be optimistic again and assume  that the singularity problem is resolved and the theory of density perturbations developed in~\cite{KOST,KKL,Khoury:2001zk} is correct. In this case one has a new problem to consider. 

Tachyonic instability, which is the source of these perturbations, amplifies not only quantum fluctuations, but also classical inhomogeneities \cite{KKL}. These inhomogeneities grow in exactly the same way as the quantum fluctuations with the same wavelength. Therefore to avoid cosmological problems {\it the initial classical inhomogeneities of the branes must be below the level of quantum fluctuations}. In other words,  the universe on the large scale must be ideally homogeneous from the very beginning. By evaluating the initial amplitude of quantum fluctuations on the scale corresponding to the observable part of the universe one finds that the branes must be parallel to each other with an accuracy better than $10^{-60}$ on a scale $10^{30}$ times greater than the distance between the branes~\cite{KKL}. 

To understand the nature of  the problem  one may compare this scenario
with inflation. Inflation removes all previously existing inhomogeneities  and simultaneously produces small density perturbations.  Meanwhile in the ekpyrotic scenario even very small initial inhomogeneities become exponentially large. Therefore instead of resolving the homogeneity problem, the ekpyrotic scenario makes this problem much worse.

Now let us assume for a moment that we were able to solve the homogeneity problem without using inflation. But we still have the flatness/entropy problem to solve.   Suppose that the universe is closed, and initially it was filled with radiation with total entropy $S$. Then its total lifetime is given by $t \sim S^{2/3} M_p^{-1}$,
after which it collapses~\cite{book}. In order to survive until the
moment $t  \sim 10^{34} M_p$, where the inhomogeneities on the scale of the present horizon are produced in the ekpyrotic scenario, the universe must
have the total entropy greater than $10^{50}$ \cite{KKL}.  Thus in order to explain why the total entropy (or the
total number of particles) in the observable part of the universe is
greater than $10^{88}$ one must assume that it was greater than $10^{50}$
from the very beginning. This is the entropy problem~\cite{book}. If the universe initially has the Planckian temperature, its
total initial mass must be greater than $10^{50} M_p$, which is the mass problem.
Also, such a universe must have very large size from the very beginning, which is the essence of the flatness problem~\cite{book}.

In comparison, in the simplest versions of chaotic inflation scenario the
homogeneity problem is solved if our part of the universe initially was
relatively homogeneous on the smallest possible scale $O(M_p^{-1})$
\cite{Chaot}. The whole universe could have originated from a  domain with
total entropy $O(1)$ and  total mass $O(M_p)$. Once this process begins,
it leads to eternal self-reproduction of the universe in all its possible
forms~\cite{Eternal,book}. Nothing like that is possible in the ekpyrotic
scenario.

Thus, the original version of the ekpyrotic scenario  does not represent a viable alternative to inflation. Independently of all other troublesome features of the ekpyrotic scenario, it  does not solve the homogeneity, flatness and entropy problems.  Apparently, the authors of this scenario realized it, because their new model, cyclic scenario \cite{cyclic}, includes {\it an infinite number of stages of inflation.} Each new stage of inflation is supposed to ensure homogeneity of the universe at the subsequent cycle. The main difference between this version of inflationary theory and the usual one is that inflation in cyclic universe occurs before the singularity rather than immediately after it. I will describe this scenario following our recent paper with Felder, Frolov and Kofman \cite{negpot}.

\section{Cyclic universe}

\subsection{Basic scenario}

In Ref. \cite{KKL} it was pointed out that avoiding inflation in the ekpyrotic scenario requires incredible fine-tuning. If one  adds a small positive constant $V_0$ to $V(Y)\sim -e^{-\alpha m Y}$, the universe at large $Y$ becomes inflationary. In the beginning  the authors of the ekpyrotic scenario claimed that they do not need this stage of `superluminal expansion' to solve all major cosmological problems, but eventually they did exactly what we suggested in \cite{KKL}: They added a small constant $V_0 \sim 10^{-120}$ (in Planck units) to $V(Y)\sim -e^{-\alpha m Y}$, which leads to inflation at large $Y$. 

\begin{figure}[h!]
\centering\leavevmode\epsfysize=5cm \epsfbox{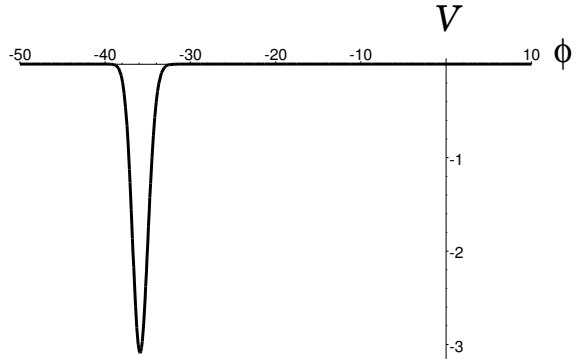} 
\caption[fig1]
{Cyclic scenario potential used in Ref. \cite{cyclic}. The potential at $\phi>1$ approaches a very small constant value $V_0 \sim 10^{-120}$. At $\phi \ll -40$ the potential vanishes.}
\label{Cyclicfig}
\end{figure}

Since the  cyclic scenario is formulated mainly in terms of the effective 4D theory with the brane separation $Y$  represented by a scalar field $\phi$, we will follow the same route. In this language, cyclic scenario, unlike ekpyrotic scenario, assumes, in accordance with \cite{KKL}, that the potential $V(\phi)$ at large $\phi$ behaves as $ V_0(1-e^{- c \phi})$ \cite{cyclic}:
\begin{eqnarray}\label{realpot}
V(\phi) =  V_0~(1-e^{-c\phi})~ F(\phi) \ .
\end{eqnarray}
In the particular example studied in the last paper of
Ref. \cite{cyclic} one has $F(\phi) = e^{-e^{-\gamma \phi}}$, $V_0 =
10^{-120}$, $c = 10$, $\gamma \approx 1/8$. This potential is shown in
Fig.  \ref{Cyclicfig}. At $\phi =0$ this potential vanishes. It
approaches its asymptotic value $V_0 = 10^{-120}$ at $\phi \gtrsim
1$. 

Inflation in this scenario is possible at $\phi \gtrsim 1$.  The potential
has a minimum at $\phi \approx -36$; the value of the potential in
this minimum is $V_{\rm min} \approx -3$ in units of Planck density \cite{cyclic}. 

Let us try to understand the origin of the parameters $c = 10$,
$\gamma \approx 1/8$ and  $V_{\rm min} \sim 3$ used in \cite{cyclic}.  According to
\cite{Khoury:2001zk}, the amplitude of density perturbations can be estimated as ${\delta_H} \lesssim 10^{-7}\xi^4 v_0^{-2/3}\sqrt {|V_{\rm min}|}$. Here $\xi \ll 1$ and $v_0$ is a ratio of Calabi-Yau volume to $M_{GUT}^{-6}$. That is why in order to obtain ${\delta_H} \sim 2\times 10^{-5}$ for $v_0 = O(1)$ \cite{Banks:1996rr}
 one should have $|V_{\rm min}| \gg 1$. The spectrum of density perturbations obtained in \cite{Khoury:2001zk}
is not blue, as in \cite{KOST}, but red, like in the simplest versions of chaotic
inflation. The spectral index is $n \approx 1-4/c^2$. Observational
data suggest that $n = 1.03 \pm 0.06$~\cite{Bond}, which implies that $c \gtrsim
10$. If one takes $c \gg 10$, and $V_{\rm min} >1$, one finds that the curvature
of the effective potential in its minimum becomes much greater than 1, i.e. the scalar particles there have mass that is much greater than $M_p$.

Once one takes $ V \sim -3$ in the minimum of the potential with $c =
10$ \cite{cyclic}, the parameter $\gamma$ can be determined
numerically: $\gamma = 0.1226$.
It would be  hard to provide explanation of the numerical value
of this parameter. If one takes, for example, $\gamma = 1/8=
0.125$, one finds $ V \sim -3\times 10^{-3}$ in the minimum of the
potential. This would reduce ${\delta_H}$ by a factor of
30. Thus, in order to have density perturbations with a correct
magnitude one should fine-tune the value of $\gamma = 0.1226$ with
accuracy better than 1\%.

According to cyclic scenario, right now the scalar field is large, $\phi > 1$, so that $V(\phi) \approx V_0 \sim 10^{-120}$. This corresponds to the present stage of accelerated expansion of the universe. Gradually the field $\phi$ begins drifting to the left and falls towards the minimum of $V(\phi)$. On the way there the curvature of the potential becomes negative, and therefore small perturbations of the field $\phi$ are generated by the mechanism explained in \cite{tach,KOST,KKL}. These are the fluctuations that are supposed to produce density perturbations after the singularity.

Then the field reaches the region with $V(\phi) <0$. At some point the total energy (including kinetic energy) vanishes. At that time,  according to the Friedmann equation for a flat universe ${\dot a^2/a^2} = \rho/3$, the universe stops expanding and begins to contract to a singularity. Collapse leads to  negative  friction coefficient $3H$ in the term $3H\dot\phi$, which accelerates the evolution of the field $\phi$.  A numerical
investigation of the motion of the field moving from $\phi >0$
in a theory with this potential shows that its kinetic energy at the
moment when $\phi$ reaches the minimum of the effective potential is
$O(10^2)M_p^4$.   When the field approaches $\phi \sim -39$, where the
effective potential becomes flat, the kinetic energy of the field
$\phi$ becomes $\sim 10^6 M_p^4$, i.e. a million times greater than the
Planck density!

At  this stage a weaker soul could falter, but we must proceed because
we did not describe the complete scenario yet.

The subsequent evolution develops very fast. When the field accelerates enough it enters the regime $\dot\phi^2 \gg
V(\phi)$ and continues moving to $-\infty$ with a speed practically independent
of $V(\phi)$: $\phi \sim \ln t$, $\dot\phi \sim t^{-1}$, where $t$ is the time remaining until the Big Crunch singularity.  For all potentials $V(\phi)$ growing at large
$\phi$ no faster than some power of $\phi$ one has $\dot\phi^2/2$
growing much faster than $V(\phi)$ (a power law singularity versus a
logarithmic singularity). This means that one can  neglect
$V(\phi)$ in the investigation of the singularity, virtually
independently of the choice of the potential at large $|\phi|$ \cite{negpot}.  This regime corresponds to a `stiff' equation of state $p = \rho$.

Usually, the Big Crunch singularity is considered the end of the evolution of the
universe. However, in the cyclic scenario it is assumed that the
universe goes through the singularity and re-appears again. When
it appears, in the first approximation it looks exactly as it was
before, and the scalar field moves back exactly by the same trajectory
by which it reached the singularity \cite{Seiberg}.

This is not a desirable cyclic regime. Therefore it is assumed in
\cite{cyclic} that the value of kinetic energy of the field $\phi$
{\it increases} after the bounce from the singularity. This increase
of energy of the scalar field is supposed to appear as a result of particle production at the moment
of the brane collision (even though one could argue that usually
particle production leads to an opposite effect). It is very hard to verify the validity of this crucial assumption since the consistency of the 5D picture proposed in \cite{cyclic} is questionable, see e.g. \cite{Rasanen:2001hf}. But let us just assume that this is indeed the case because otherwise the whole scenario does not work and we have nothing to discuss. If the increase of
the kinetic energy is large enough, the field $\phi$ rapidly rolls
over the minimum of $V(\phi)$ in a state with a positive total energy
density, and continues its motion towards $\phi > 0$.  The kinetic energy
of the field decreases faster than the energy of matter produced at
the singularity.  At some moment the energy of matter begins to
dominate. Eventually  the energy density of ordinary matter becomes smaller than
$V(\phi)$ and the present stage of inflation (acceleration of the
universe) starts again. This happens if the field $\phi$ initially moved fast enough to reach the plateau of the effective potential at $\phi \gg 1$. 

Because the potential at $\phi \gg 1$ is very flat, the field may stay there
for a long time and inflation will make the universe flat and empty. Eventually the field rolls towards the minimum of the potential again, and the universe enters a new cycle of contraction and expansion.

As we see, this version of the ekpyrotic scenario is not an
alternative to inflation anymore. Rather it is a very specific version
of inflationary theory. The major cosmological problems are supposed
to be solved due to exponential expansion in a vacuum-like state, i.e. by inflation, even
though it is a low-energy inflation and the mechanism of production of density perturbations in this scenario is non-standard. Let us remember that Guth's first paper on
inflation \cite{Guth} was greeted with so much enthusiasm precisely
because it proposed a solution to the homogeneity, isotropy, flatness
and horizon problems due to exponential expansion in a vacuum-like state, 
even though it didn't address the formation of
large scale structure. The Starobinsky model that was proposed a year
earlier \cite{Star} could account for large scale structure
and the observed CMB anisotropy \cite{Mukh}, but it did not
attract as much attention because it did not address the possibility
of solving these initial condition problems.

In fact, the stage of acceleration of the universe in the cyclic model
is {\it eternal inflation}. Eternal inflation   occurs if $V'^2 \ll V^3$ \cite{Eternal,LLM}. For the potential $V(\phi)$ used in
the cyclic model this condition is satisfied at large $\phi$ since  $V = const$ in the limit $\phi \to
\infty$, whereas $V' \to 0$ in this limit. Thus the universe
at large $\phi$ (at $\phi > 15$ in the model of Ref. \cite{cyclic}) enters the stage of eternal self-reproduction, quite
independently of the possibility to go through the singularity and
re-appear again. In other words, the universe in the cyclic scenario
is not merely a chain of eternal repetition, as expected in \cite{cyclic}, but a growing
self-reproducing inflationary fractal of the type discussed in
\cite{Vilenkin:xq,Eternal,LLM}.

One may wonder, however, whether this version of inflationary theory
is good enough to solve all major cosmological problems. Indeed,
inflation in this scenario may occur only at a density 120 orders of
magnitude smaller than the Planck density. If, for example, one
considers a closed universe filled with matter and a scalar field with
the potential used in the cyclic model, it will typically collapse
within the Planck time $t \sim 1$, so it will not survive until the
beginning of inflation in this model at $t \sim 10^{60}$.  For
consistency of this scenario, the overall size of the universe at the
Planck time must be greater than $l \sim 10^{30}$ in Planck units,
which constitutes the usual flatness problem. The total entropy of 
a hot universe that may survive until
the beginning of inflation at $V \sim 10^{-120}$ should be greater
than $10^{90}$, which is the entropy problem \cite{book}.  An estimate
of the probability of quantum creation of such a universe ``from
nothing'' gives $P \sim e^{-|S|} \sim \exp\left(-{ 24\pi\over V_0
}\right) \sim e^{-120}$ \cite{Creation}.

There are many other unsolved problems related to this theory, such as
the origin of the potential $V(\phi)$ \cite{KKL} and the 5D
description of the process of brane motion and collision
\cite{KKLT,Rasanen:2001hf}. In particular, the cyclic scenario assumes
that the distance between the branes is not stabilized, i.e. the field $\phi$ at present is nearly massless. This may lead to a strong violation of the equivalence principle. This is one of the main reasons why it is usually assumed that
the branes in Ho\u{r}ava-Witten theory must be stabilized. To avoid this problem, the authors of \cite{cyclic} introduce the function $\beta(\phi)$ describing interaction of the scalar field with matter. The violation of the equivalence principle can be avoided if $(\ln \beta(\phi))_{,\phi} \ll 10^{-3}$. However, in the Kaluza-Klein limit, in which the 4D approximation used in \cite{cyclic} could be valid, one has $\beta(\phi) \sim e^{\phi/\sqrt 6}$ \cite{cyclic}. In this approximation, one has  $(\ln \beta(\phi))_{,\phi} = 1/\sqrt 6$, and the equivalence principle is strongly violated.

We will not discuss these problems here any longer. Instead of that, we will
concentrate on the phenomenological description of possible cycles
using the effective 4D description of this scenario. This will allow
us  to find out whether the cyclic regime is indeed a natural feature
of the scenario proposed in \cite{cyclic}.

\subsection{Are there any cycles in the cyclic scenario?}

As we have seen, the existence of the cyclic regime requires investigation of particle production in the singularity. Fortunately, this subject was intensely studied more than 20 years ago.
The main results can be summarised as follows. Since $H \sim t^{-1}$ in the standard big bang theory, the
curvature scalar $R$ in the universe dominated by a kinetic energy of a scalar field  behaves as $t^{-2}$ . Scalar particles
minimally coupled to gravity, as well as gravitons and helicity 1/2
gravitinos \cite{Kallosh:2000ve}, are not conformally invariant; their
frequencies thus experience rapid nonadiabatic changes induced by the
changing curvature. These changes lead to particle production due to
nonadiabaticity with typical momenta $k^2 \sim R \sim t^{-2}$.  The
total energy-momentum tensor of such particles produced at a time $t$
after (or before) the singularity is $T_{\mu\nu} \sim O(k^4) \sim R^2
\sim t^{-4}$ \cite{ford,grib}. Comparing the density of
produced particles with the classical matter or radiation density of
the universe $\rho \sim t^{-2}$, one finds that the density of created
particles produced at the Planck time $t \sim 1$ is of the same order
as the total energy density in the universe. Moreover, if one  bravely attempts to describe the situation near the singularity, at $R > 1$, then one may conclude that the contribution of particles produced near the singularity (as well as the contribution of quantum corrections to $T_{\mu\nu}$) is always greater than the energy momentum tensor of the classical scalar field.   

This argument will be very important for us. The
existence of even a small amount of particles created near the singularity may have a significant effect on the motion of the field. Indeed, the kinetic energy of the
scalar field $\dot\phi^2/2$ in the regime $\dot\phi^2/2 \gg V(\phi)$
decreases as $a^{-6}$. Meanwhile, the density of ultrarelativistic particles decreases
as $a^{-4}$. Therefore at some moment $t_0$ the energy density of ultrarelativistic particles eventually becomes greater than $\dot\phi^2/2$. In this regime (and neglecting $V(\phi)$) one can show that  $\dot\phi  = \dot\phi_0 {a_0^3\over
 a^3} = \dot\phi_0 \left({t_0\over t}\right)^{3/2}$.
Even if this regime continues for
an indefinitely long time, the total change of the field $\phi$ during this
time remains quite limited.  Indeed,
\begin{equation} \label{changephi}
\Delta\phi \leq \int\limits_{t_0}^\infty \dot\phi dt = \dot\phi_0
\int\limits_{t_0}^\infty \left({t_0\over t}\right)^{3\over 2}
dt = 2~\dot\phi_0 t_0 .
\end{equation}
If $t_0$ is the very beginning of radiation domination ($\dot\phi_0^2/2
\sim \rho_{\rm total}$), then $H_0 \sim t_0^{-1}  \sim \dot\phi_0$. Therefore
\begin{equation}\label{rchange}
\Delta\phi \lesssim 1   
\end{equation}
in Planck units (i.e. $\Delta\phi \lesssim M_p$).   

This simple result has several important implications.  In particular, if the
motion of the field in a matter-dominated universe begins at $|\phi|
\gg 1$, then it can move only by $\Delta\phi \lesssim 1$. Therefore in
theories with flat potentials the field always remains frozen at
$|\phi| \gg 1$.
It begins moving again only when the Hubble constant decreases
and $|3H\dot\phi|$ becomes comparable to $|V,_\phi|$. But in this case
the condition $ 3H\dot\phi \approx |V,_\phi|$ automatically leads to
inflation in such theories   as $m^2\phi^2/2$ for   $\phi
\gg 1$.
This means that even a small amount of matter or radiation may
increase the chances of reaching a stage of inflation, see   \cite{Toporensky:1999pk,negpot}. 

In application to the cyclic scenario this result implies that   the scalar field in presence of ultrarelativistic matter created near the singularity should immediately loose its kinetic energy and freeze to the left of the minimum of the effective potential at  Fig. \ref{Cyclicfig}. Then it slowly falls to the minimum. If the field would move rapidly, as expected in \cite{cyclic} , it would roll to positive $\phi$ without triggering the collapse of the universe. But a slow motion leads  to a  collapse of the universe \cite{negpot}, instead of the inflationary stage anticipated in \cite{cyclic}.

But what if we are wrong, and for some reason particle production near the singularity was inefficient \cite{Tolley:2002cv}? Independently of it, a later stage of efficient particle production is unavoidable.
Indeed, at $\phi <-40$ the mass squared of the scalar field  $m^2 = V''$ vanishes. Near the minimum one has $m  \gtrsim O(1)$ in Planck units, and at $\phi > -32$ the mass almost exactly vanishes again. The transition from $m \ll 1$ to $m\sim 1$ and back to $m \ll 1$ occurs within the time $\Delta t \sim 1$. Thus, the change of the mass was strongly non-adiabatic, with typical frequency $\sim 1$. This should lead to production of ultrarelativistic particles $\phi$ with Planckian density \cite{KLS}. These particles, in their turn, should immediately freeze the motion of the field $\phi$ near $\phi \sim -30$. The field never reaches the region $\phi > 1$, and inflation never begins. Instead of that, the field again slowly falls to the minimum of $V(\phi)$ and the universe collapses.

Thus, unless one makes some substantial modification to the scenario proposed in \cite{cyclic}, it does not work in the way anticipated by its authors.

\subsection{\label{bicycling}Cycles and epicycles}

Of course, one could save this scenario by adding new epicycles to it. For example, if particle production near the singularity at $\phi = -\infty$ freezes the field to the left of the minimum  of $V(\phi)$,  then the universe collapses for the second time when the field runs to $\phi = +\infty$. If the field bounces back from the singularity  at $\phi = +\infty$ and then   freezes  at $\phi >0$ due to efficient particle production at the singularity,  a stage of low-energy inflation begins.   This scenario is quite different from the one proposed in \cite{cyclic}, but it might work. Density perturbations in this scenario will be produced due to tachyonic instability in the potential with the shape determined by the term $e^{-e^{-\gamma \phi}}$.

Also, one may consider
effects related to non-relativistic particles produced at the
singularity. These particles contribute to the equation of motion for
the field $\phi$ by effectively increasing its potential energy
density \cite{cyclic}. They may push the field towards positive values
of the field $\phi$ despite the effects described above. However, this
would add an additional complicated feature to a scenario that is already quite
speculative. 

\begin{figure}[h!]
\centering\leavevmode\epsfysize=5cm \epsfbox{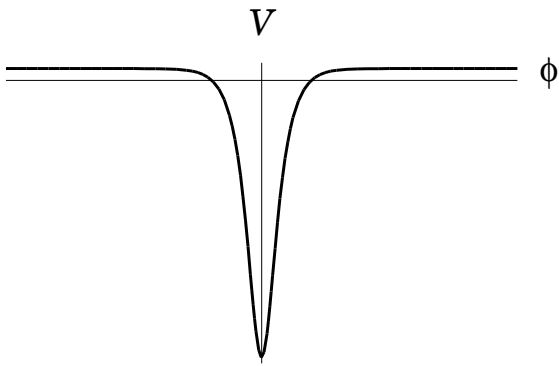} 
\caption[fig1]
{Symmetric scalar field potential in the bicycling
scenario. At large values of $|\phi|$ one has $V(\phi) \approx V_0
\sim 10^{-120}$ and there is a minimum at $\phi = \phi_0$.}
\label{Cyclic3}
\end{figure}

But one can do something simpler. For
example, instead of the asymmetric potential shown in
Fig. \ref{Cyclicfig}, one may consider a symmetric
potential, as in Fig. \ref{Cyclic3}. As a toy model,  one may consider, e.g., a potential $V(\phi)  =  V_0(1 -A \cosh^{-1}(\phi - \phi_0))$, where $A>1$  and $\phi_0$ are some constants. 

In the beginning, the scalar field is large and positive and it slowly
moves towards the minimum. When it falls to the minimum the universe
begins to contract and the field is rapidly accelerated towards the
singularity at $\phi = -\infty$. As we already mentioned, the
structure of the singularity is not sensitive to the existence of the
potential, especially if it is as small as $V_0 \sim
10^{-120}$. 

Now let us assume, as in \cite{cyclic}, that the field $\phi$ bounces
from the singularity and moves back. The kinetic energy of the field
rapidly drops down because of radiation, so it freezes at the plateau to the left of the minimum of $V(\phi)$.   At this
stage the energy density is dominated by particles produced near the
singularity. Then the universe cools down while
the field is still large and negative and the late-time stage of
inflation begins. During this stage the field slowly slides towards
the minimum of the effective potential and then rolls towards the
singularity at $\phi \to \infty$. When it bounces from the
singularity, a new stage of inflation begins. The universe in this
scenario enters a cyclic regime with twice as many cycles as in the
original cyclic scenario of Ref. \cite{cyclic}. The same regime will appear even if the potential is asymmetric but positive both at $\phi \ll 0$ and at $\phi \gg 0$. We have  called it the {\it bicycling scenario} \cite{negpot}.

An advantage of this scenario is that it may work even if, as we expect, a lot of
radiation is produced at the singularity and the field $\phi$ rapidly
loses its kinetic energy. However, if in order to have density
perturbations of a sufficiently large magnitude one needs to have a
potential with a super-Planckian depth $V(\phi) <-1$, as in
\cite{Khoury:2001zk,cyclic}, then this scenario has the same problem
as the scenario considered in the previous section. The kinetic energy
of the field $\phi$ becomes greater than the Planck density as soon as
it rolls to the minimum of $V(\phi)$. It becomes even much greater
when the field rolls out of the minimum, and the 4D description fails.

Thus, bicycling scenario is  more robust  than the original version of the cyclic scenario \cite{cyclic}, but it is still very problematic.

\subsection{\label{cyclicinflation}Cycles with inflationary density
perturbations}

Even though it may not be easy to solve the singularity problem, the 
old idea that the big bang is not the beginning of the universe but a
point of a phase transition is quite interesting, see
e.g. \cite{Tolman-1931}-\cite{PBB}. However, the more assumptions
about the singularity one needs to make, the less trustworthy are the
conclusions. In this respect, inflationary theory provides us with a
unique possibility to construct a theory largely independent of any
assumptions about the initial singularity. According to this theory,
the structure of the observable part of the universe is determined by
processes at the last stages of inflation, at densities much smaller
than the Planck density. As a result, observational data practically
do not depend on the unknown initial conditions in the early universe.

But this very fact puts all theories of passing through the singularity in a difficult position. Indeed, one could argue that if the universe passes through the singularity, but afterwards there is a stage of inflation that erases all observational consequences of this event, then why bother? That is why some authors were trying to avoid having inflation after the singularity.

In my opinion, the possibility that the universe may pass  through the cosmological singularity is extremely interesting independently of absence or presence of any  observational consequences of this event. We need to have a consistent cosmological theory, including all possible stages of the evolution of the universe, whether they are accessible to observations or not.

An interesting attempt to construct such a theory  was made in the context of the Pre-Big Bang scenario \cite{PBB}. The main idea of this theory is extremely attractive. However, the authors of the PBB scenario decided to consider only those versions of the theory that do not have inflation after the singularity. As a result, in addition to the unsolved singularity problem, the PBB theory  has a problem with production of adiabatic density perturbations with flat spectrum. More importantly, without help of inflation the PBB theory does not solve the homogeneity, isotropy, flatness and entropy problems \cite{Kaloper}. So why do not we add a stage of inflation after the singularity? Of course, this would make the passage through the singularity irrelevant from the point of view of observations. However, this   would solve the major cosmological problems (if the singularity problem is solved) and  make the whole scenario, including the non-trivial stage of the pre-big bang evolution, consistent with observational data.

The situation with the cyclic scenario is very similar. I believe that the necessity to solve the singularity problem and to know exactly what happens with small density perturbations at the singularity is a significant step back from the simple picture provided by the usual versions of inflationary theory.  
Since the cyclic scenario does require repeated periods of inflation anyway, it would be nice to avoid the vulnerability of this scenario with respect to the unknown
physics at the singularity by placing the stage of inflation  before 
the stage of large scale structure formation rather than after it.

\begin{figure}[h!]
\centering\leavevmode\epsfysize=5cm \epsfbox{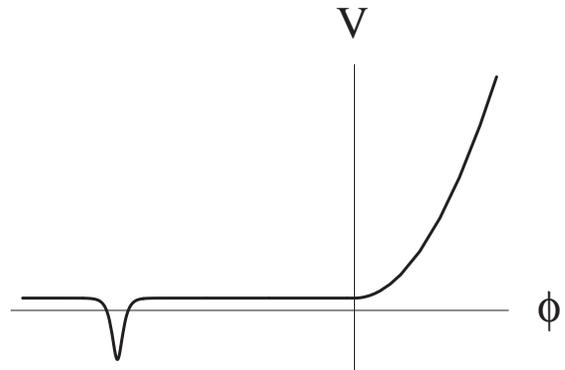} 
\caption[fig1]
{Scalar field potential in the cyclic scenario incorporating a
stage of chaotic inflation. The potential has a minimum at $\phi_0 <0$. Inflationary perturbations are generated and the large-scale structure of the universe is produced  at $\phi > 0$. }
\label{Cyclic5}
\end{figure}

In order to do it \cite{negpot}, one may consider  a toy model with a potential 
\begin{eqnarray}
V(\phi) &=& V_0(1 -A \cosh^{-1}(\phi - \phi_0)) \qquad     {\rm for }~~ \phi<0 \ , \\
V(\phi) &=& V_0(1 -A \cosh^{-1}(\phi - \phi_0)) + {m^2\over 2}\phi^2 ~~  {\rm for }~~ \phi> 0 \ .\nonumber
\end{eqnarray}
Thus, for $m=0$ this is the same potential as in the bicycling scenario proposed in the previous section.  If one takes  $\phi_0 \ll -1$, then the potential at $\phi < 0$ looks  very similar to the potential of the original cyclic model, see Fig. \ref{Cyclicfig}, but it is positive everywhere except a small vicinity of $\phi_0$, see Fig. \ref{Cyclic5}. Also, we will not need to have a very deep minimum of the effective potential because density perturbations will be produced by the standard inflationary mechanism. Therefore one can take $A$ just a bit greater than 1. At $\phi >0$ the potential coincides with the simplest chaotic inflation potential $V_0 + {m^2\over 2}\phi^2$ considered in the beginning of this paper.

Now let us assume that initially the universe was slowly inflating in a state with $\phi < \phi_0$. Then the scalar field started moving towards the minimum of the effective potential, as in the cyclic scenario (though in an opposite direction). When the field approaches the minimum, the universe begins to collapse. After that moment, the field begins growing with an increasing speed. Let us assume that one can ignore  the effective potential ${m^2\over 2}\phi^2$ that appears only at $\phi > 0$.  According to the investigation performed in \cite{negpot}, for $A = O(1)$ the kinetic energy of the field $\phi$ will reach the Planck value $\dot\phi^2/2 \sim 1$ at $\phi-\phi_0 \sim 100$.  Meanwhile, for $m =3 \times 10^{-6}$ (COBE normalization), the effective potential ${m^2\over 2}\phi^2$ reaches the Planck value only at $\phi  \sim 10^5$, where the kinetic energy of the field $\phi$ would be much greater than 1. Thus, if one takes $A = O(1)$, $\phi_0 \sim 100$,  one can completely ignore $V(\phi)$ during the investigation of the development of the singularity, as well as during the first stages of the motion of the field $\phi$ bounced back from $+\infty$.

However, according to our discussion of particle production and their effect on the motion of the scalar field, one may expect that after bouncing from the singularity the scalar field immediately freezes and does not move (or moves relatively slowly) until its potential energy begins to dominate.   But this creates ideal initial conditions for the beginning of a long stage of chaotic inflation! 

Of course, we do not want to pretend that we really know what happens at super-Planckian densities. Fortunately, we do not need to have an exact knowledge of the processes near the singularity. The only thing that matters for us is the assumption that quantum effects and particle production slow down the motion of the scalar field bouncing from the singularity, and  this leads to inflation.  An important advantage of this scenario is that the density perturbations produced at the end of the inflationary stage have desirable magnitude $\delta_H  \sim 2\times 10^{-5}$  independently of any assumptions about the behavior of
perturbations passing through the singularity. Also, one no longer
needs to trust calculations of perturbations at $|V(\phi)| > 1$.

Chaotic inflation ends at $\phi \sim 1$. At that time the kinetic energy of the scalar field is $\dot\phi^2/2 \sim m^2$ \cite{book}. The energy density of particles produced at the singularity vanishes during inflation. However, new particles with energy density $H^4 \sim m^4$ are produced at the end of inflation because of the gravitational effects \cite{ford,grib,QuintInfl} and because of the nonadiabatic change of the mass of the field $\phi$. These particles in  their turn freeze the rolling field $\phi$. This field freezes even more efficiently if it interacts with any other particles and give them mass $\sim g|\phi|$. In this case the mechanism of instant preheating leads to production of these particles, which freezes the motion of the field $\phi$ at $\phi \sim - \ln g^{-2}$, \cite{Instant,NO}.
The  new particles created  at that stage and the products of their interactions constitute the
matter contents of the observable universe.

After many billions of years the density of ordinary matter decreases, and the energy
density of the universe becomes determined by $V(\phi) \approx
V_0$. The universe enters the present  stage of low energy inflation. This stage lasts for a very long time. During this time the field $\phi$  slowly rolls towards
 $\phi = \phi_0$. Then it falls to the minimum, runs to $-\infty$, bounces back after the singularity, slows
down due to radiation, experiences low-energy inflation at $\phi < \phi_0$,  rolls
down to the minimum of $V(\phi)$ again, runs to $+\infty$, bounces back, and a new stage of chaotic inflation begins.

\begin{figure}[h!]
\centering\leavevmode\epsfysize=5cm \epsfbox{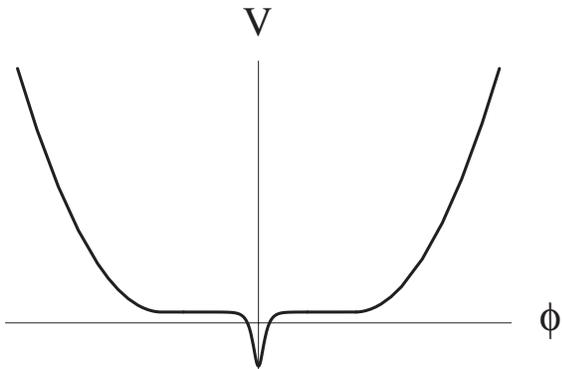} 
\caption[fig1]
{Scalar field potential in the cyclic scenario incorporating a
stage of chaotic inflation. Inflationary perturbations are generated and the large-scale structure of the universe is produced both at $\phi < 0$ and at $\phi > 0$.}
\label{Cyclic6}
\end{figure}

In this model
inflationary perturbations are generated only every second time after
the universe passes the singularity (at $\phi <0$, but not at $\phi
<0$). The model can be further extended by making the potential rise
both at $\phi \to \infty$ and at $\phi \to -\infty$, see
Fig. \ref{Cyclic6}. In this case the stage of high-energy inflation
and large-scale structure formation occurs each time after the
universe goes through the singularity.

 Thus we see that it is possible to propose a scenario describing an
oscillating inflationary universe without making any assumptions about
the behavior of non-inflationary perturbations near the
singularity \cite{negpot}. Another important advantage of this scenario is that
inflationary cycles may begin in a universe with initial size as small
as $O(1)$ in units of the Planck length, just as in the standard
chaotic scenario \cite{Chaot}.

Even though this scenario is free of many problems that plagued the old cyclic scenario \cite{cyclic}, it still remains  complicated and speculative. 
The main problem of this model is that one still must assume that
somehow the universe can go through the singularity. However, now this
assumption is no longer required for the success of the scenario since
the large scale structure of the observable part of the universe in this scenario does not depend on processes near the singularity. This allows us to remove the
remaining epicycles of this model. 

 \begin{figure}[h!]
\centering\leavevmode\epsfysize=5cm \epsfbox{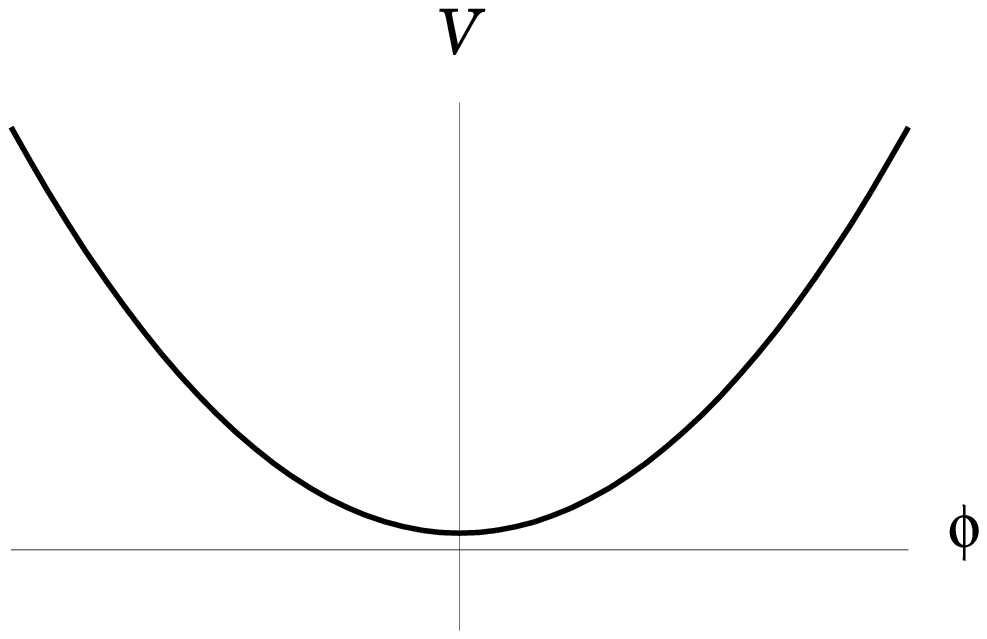} 
\caption[fig1]
{The scalar field potential that appears after the step-by-step simplification of the cyclic scenario. It coincides with the simplest version of chaotic inflation scenario, Fig. \ref{Cyclic7}. }
\label{Cyclic7a}
\end{figure}

Indeed, the main source of all the
problems in this model is the existence of the minimum of the
effective potential with $V(\phi)<0$. Once one cuts this minimum off,
the potential becomes extremely simple, see Fig. \ref{Cyclic7a}, and
all problems mentioned above disappear. In particular, one may use the
simplest harmonic oscillator potential ${m^2\over 2}\phi^2 +V_0$ 
with $V_0 \sim 10^{-120}$
considered in the beginning of our paper. This theory describes an
eternally self-reproducing chaotic inflationary universe, as well as
the late stage of accelerated expansion (inflation) of the universe
driven by the vacuum energy $V_0>0$.

\section{Conclusions}
   
In this paper I have briefly reviewed the basic principles of
inflationary cosmology. During the last 20 years ago this theory has
changed considerably and gradually become the standard framework for
the investigation of the early universe. Recent observational data has
brought us additional reasons to believe that we might be on the right
track.  It is quite encouraging that so far the simplest versions of
inflationary cosmology seem to be in good agreement with
observations. Still there are many things to do. We do not know which
version of inflationary theory is the best.  We do not even know
whether the inflaton field is a scalar field, as in old, new and
chaotic inflation, or something related to the curvature scalar, as in
the Starobinsky model, or something else entirely like the logarithm
of the radius of compactification or a distance between branes. It is
possible to have several different stages of inflation; one could
solve the homogeneity and isotropy problems while another could
produce density perturbations. This latter stage may look like
exponential expansion in all directions, or it may be viewed as
exponential expansion on some particular hypersurface in a higher
dimensional space.

Thus, there exist many different versions of inflationary cosmology,
and many new ones will certainly appear with every new development in
the theory of fundamental interactions.  But one may wonder whether
these new developments will eventually allow us to find a consistent
non-inflationary cosmological theory?  While we cannot give a general
answer to this question, we hope that our investigation of the
ekpyrotic/cyclic scenario demonstrates how difficult it is to
construct a consistent cosmological theory without using inflation.

The original version of the ekpyrotic scenario \cite{KOST} contained
many incorrect and unproven statements  \cite{KKL,KKLT}.  In particular, instead of expansion of the colliding branes described in \cite{KOST}, one has
contraction to a singularity \cite{KKLT,Khoury:2001bz}. Despite the
optimistic statements of the authors of \cite{KOST}, the singularity
problem in this scenario remains unsolved. The theory of density
perturbations in this scenario is controversial; most authors believe
that the mechanism of generation of density perturbations proposed in
\cite{KOST,Khoury:2001bz} does not lead to adiabatic perturbations
with a flat spectrum
\cite{Lyth:2001pf,Brandenberger:2001bs,Hwang:2002ks,Martin:2001ue}. Most
importantly, this scenario offers no solution to major cosmological
problems such as the homogeneity, flatness and entropy problems. In
fact, the homogeneity problem in this scenario is even much more
complicated than in the usual non-inflationary big bang theory
\cite{KKL}.

As for the cyclic scenario \cite{cyclic}, its authors recently issued
a paper advertising this scenario in the popular press
\cite{cyclicnew} and another one aimed at astrophysicists
\cite{cyclicastro}. These new papers, which were supposed to give a
summary of the state of the cyclic universe theory, omitted any
mention of the criticisms of the ekpyrotic/cyclic scenario contained
in \cite{negpot} and in
\cite{KKL,KKLT,Lyth:2001pf,Brandenberger:2001bs,Hwang:2002ks,Martin:2001ue,Rasanen:2001hf}.
It was claimed in \cite{cyclicastro} that the cyclic scenario
{``\it is able to reproduce all of the successful predictions of the
consensus model\, {\rm (i.e. of inflationary cosmology -A.L.)} with
the same exquisite detail.''}  They continued by saying that ``{\it
All of the differences between the two paradigms harken back to the
disparate assumptions about whether there is a ``beginning'' or
not.}'' Then they said that ``{\it if the big bang were not a
beginning, but, rather, a transition from a pre-existing contracting
phase, then the inflationary mechanism would fail.}''

I cannot agree with these statements, for the reasons explained
in Ref. \cite{negpot} and in this paper. First of all, the cyclic
scenario uses an infinite number of stages of inflation to solve the
homogeneity problem, and therefore it is not an alternative to
inflation. These stages of inflation are possible only if the universe
is always huge. Thus, the flatness and entropy problems remain
unsolved. This scenario is plagued by the singularity problem, and the
situation with density perturbations is as obscure as in the ekpyrotic
scenario. Finally, as was shown in \cite{negpot},
the simplest version of this scenario does
not work in the way anticipated in \cite{cyclic} because of effects
related to particle production. Just as in the ekpyrotic scenario
\cite{KOST}, where the universe collapses instead of expanding
\cite{KKLT}, in the cyclic model of Ref. \cite{cyclic} the universe
collapses instead of inflating \cite{negpot}.

In this paper, following \cite{negpot}, I described several ways to
modify the cyclic model of Ref. \cite{cyclic} in order to avoid this
problem.  The simplest way is to make the potential  
positive both at $\phi \gg 0$ and at $\phi \ll 0$. This doubles the
number of inflationary cycles but still leaves the scenario vulnerable
with respect to the unknown physics at the singularity and to the
problem of density perturbations.

One can resolve these problems by adding a stage of chaotic inflation
after the singularity \cite{negpot}. Indeed, contrary to the statement
of Ref. \cite{cyclicastro}, there is no reason to believe that the
existence of a stage of contraction prior to the singularity should
disallow inflation after the singularity. One should not represent
inflation and the existence of the universe prior to the singularity
as two incompatible possibilities.

The addition of an inflationary stage after the singularity allows one
to use the standard inflationary mechanism of generation of density
perturbations and makes all observational consequences of the theory
indistinguishable from those in the usual chaotic inflation.  However,
in this scenario one still needs to assume that the universe can pass
through the singularity, and one should use a non-standard mechanism
of reheating after inflation. 

But once we agree that one needs
inflation in one way or another to solve the major cosmological
problems, then why should one suffer with such complicated versions of
inflationary theory? All of the problems discussed above completely
disappear if one removes the minimum of the potential with $V(\phi)
<0$. This final simplification reduces the theory to the standard
chaotic inflation scenario describing an eternally self-reproducing
inflationary universe, as well as the late stage of accelerated
expansion of the universe.

This paper is dedicated to Stephen Hawking celebrating his 60th
birthday, and therefore I would like to finish it on a positive
note. So here it is:
\begin{center}
{\bf The ekpyrotic/cyclic scenario is the best alternative to
inflation that I am aware of.}
\end{center}

I really mean it. I think that we should be very grateful to its
authors. Indeed, if a model speculating about an infinite number of
inflationary stages separated by an infinite number of singularities
is the best alternative to inflation invented during the last 20
years, this means that inflationary theory is in a very good shape.

\

{\bf Note Added:} After I  wrote this paper, a new alternative to inflation was proposed by Hollands and  Wald \cite{Hollands:2002yb}. The authors admit  that their new proposal does not solve any of the major cosmological problems addressed by inflation. However, they proposed a new hypothetical mechanism of generation of density perturbations with flat spectrum. We will discuss details of their proposal in a separate publication \cite{alt}. For the purposes of this paper it is sufficient to say that according to this new proposal the perturbations on the scale of the present horizon were generated at density $10^{96}$ times greater than the Planck density. This shows again how difficult it is to construct a consistent cosmological theory without using inflation.



\begin{thebibliography}{99}\label{bib} 



\bibitem{New} 
A.~D.~Linde,
``A New Inflationary Universe Scenario: A Possible Solution Of The Horizon, Flatness, Homogeneity, Isotropy And Primordial Monopole Problems,''
Phys.\ Lett.\ B {\bf 108}, 389 (1982).
 





\bibitem{HMS}  
S.~W.~Hawking, I.~G.~Moss and J.~M.~Stewart,
``Bubble Collisions In The Very Early Universe,''
Phys.\ Rev.\ D {\bf 26}, 2681 (1982).
 


\bibitem{Guth} A.~H.~Guth,
``The Inflationary Universe: A Possible Solution To The Horizon And Flatness Problems,''
Phys.\ Rev.\ D {\bf 23}, 347 (1981).
 

\bibitem{GuthWeinb}
A.~H.~Guth and E.~J.~Weinberg,
``Could The Universe Have Recovered From A Slow First Order Phase Transition?,''
Nucl.\ Phys.\ B {\bf 212}, 321 (1983).




\bibitem{HawkMoss}
S.~W.~Hawking and I.~G.~Moss,
``Supercooled Phase Transitions In The Very Early Universe,''
Phys.\ Lett.\ B {\bf 110}, 35 (1982).
 


\bibitem{AlStein} A.~Albrecht and P.~J.~Steinhardt,
``Cosmology For Grand Unified Theories With Radiatively Induced Symmetry Breaking,''
Phys.\ Rev.\ Lett.\  {\bf 48}, 1220 (1982).
 

\bibitem{Vilenkin:wt}
A.~Vilenkin and L.~H.~Ford,
``Gravitational Effects Upon Cosmological Phase Transitions,''
Phys.\ Rev.\ D {\bf 26}, 1231 (1982).
 

\bibitem{Linde:uu}
A.~D.~Linde,
``Scalar Field Fluctuations In Expanding Universe And The New Inflationary Universe Scenario,''
Phys.\ Lett.\ B {\bf 116}, 335 (1982).
 

\bibitem{Mukh} V.~F.~Mukhanov and G.~V.~Chibisov,
``Quantum Fluctuation And `Nonsingular' Universe,''
JETP Lett.\  {\bf 33}, 532 (1981)
[Pisma Zh.\ Eksp.\ Teor.\ Fiz.\  {\bf 33}, 549 (1981)].
 



\bibitem{Hawk} S.~W.~Hawking,
``The Development Of Irregularities In A Single Bubble Inflationary Universe,''
Phys.\ Lett.\ B {\bf 115}, 295 (1982); 
A.~A.~Starobinsky,
``Dynamics Of Phase Transition In The New Inflationary Universe Scenario And Generation Of Perturbations,''
Phys.\ Lett.\ B {\bf 117}, 175 (1982); A.~H.~Guth and S.~Y.~Pi,
``Fluctuations In The New Inflationary Universe,''
Phys.\ Rev.\ Lett.\  {\bf 49}, 1110 (1982); J.~M.~Bardeen, P.~J.~Steinhardt and M.~S.~Turner, ``Spontaneous Creation Of Almost Scale - Free Density Perturbations In An Inflationary Universe,''
Phys.\ Rev.\ D {\bf 28}, 679 (1983).

\bibitem{Mukh2} V.~F.~Mukhanov,
``Gravitational Instability Of The Universe Filled With A Scalar Field,''
JETP Lett.\  {\bf 41}, 493 (1985)
[Pisma Zh.\ Eksp.\ Teor.\ Fiz.\  {\bf 41}, 402 (1985)];  V.~F.~Mukhanov, H.~A.~Feldman and R.~H.~Brandenberger,
``Theory Of Cosmological Perturbations,''
Phys.\ Rept.\  {\bf 215}, 203 (1992).



\bibitem{Kirzhnits}
D.~A.~Kirzhnits,
``Weinberg Model In The Hot Universe,''
JETP Lett.\  {\bf 15}, 529 (1972)
[Pisma Zh.\ Eksp.\ Teor.\ Fiz.\  {\bf 15}, 745 (1972)]; D.~A.~Kirzhnits and A.~D.~Linde,
``Macroscopic Consequences Of The Weinberg Model,''
Phys.\ Lett.\ B {\bf 42}, 471 (1972); S.~Weinberg,
``Gauge And Global Symmetries At High Temperature,''
Phys.\ Rev.\ D {\bf 9}, 3357 (1974); L.~Dolan and R.~Jackiw,
``Gauge Invariant Signal For Gauge Symmetry Breaking,''
Phys.\ Rev.\ D {\bf 9}, 2904 (1974);
D.~A.~Kirzhnits and A.~D.~Linde,
``A Relativistic Phase Transition,''
Sov.\ Phys.\ JETP {\bf 40}, 628 (1975)
[Zh.\ Eksp.\ Teor.\ Fiz.\  {\bf 67}, 1263 (1974)]; D.~A.~Kirzhnits and A.~D.~Linde,
``Symmetry Behavior In Gauge Theories,''
Annals Phys.\  {\bf 101}, 195 (1976).
 


\bibitem{Chaot} A.~D.~Linde,
``Chaotic Inflation,''
Phys.\ Lett.\ B {\bf 129}, 177 (1983).
 



\bibitem{book} A.D. Linde,  {\it  Particle  Physics  and
Inflationary Cosmology} (Harwood, Chur, Switzerland, 1990).


\bibitem{Star} A.~A.~Starobinsky,
``Spectrum Of Relict Gravitational Radiation And The Early State Of The  Universe,''
JETP Lett.\  {\bf 30}, 682 (1979)
[Pisma Zh.\ Eksp.\ Teor.\ Fiz.\  {\bf 30}, 719 (1979)]; A.~A.~Starobinsky,
``A New Type Of Isotropic Cosmological Models Without Singularity,''
Phys.\ Lett.\ B {\bf 91}, 99 (1980).






\bibitem{Dolgov:1982th}
A.~D.~Dolgov and A.~D.~Linde,
``Baryon Asymmetry In Inflationary Universe,''
Phys.\ Lett.\ B {\bf 116}, 329 (1982).
 

\bibitem{DL}
A.~D.~Dolgov and A.~D.~Linde,
``Baryon Asymmetry In Inflationary Universe,''
Phys.\ Lett.\ B {\bf 116}, 329 (1982); L.~F.~Abbott, E.~Farhi and M.~B.~Wise,
``Particle Production In The New Inflationary Cosmology,''
Phys.\ Lett.\ B {\bf 117}, 29 (1982).

\bibitem{KLS} L.~Kofman, A.~D.~Linde and A.~A.~Starobinsky,
``Reheating after inflation,''
Phys.\ Rev.\ Lett.\  {\bf 73}, 3195 (1994)
[arXiv:hep-th/9405187]; L.~Kofman, A.~D.~Linde and A.~A.~Starobinsky,
``Towards the theory of reheating after inflation,''
Phys.\ Rev.\ D {\bf 56}, 3258 (1997)
[arXiv:hep-ph/9704452].
 

\bibitem{tach}  G.~Felder, J.~Garcia-Bellido, P.~B.~Greene, L.~Kofman,   
A.~Linde and I.~Tkachev,   
 ``Dynamics of symmetry breaking and tachyonic preheating,'' Phys. Rev. Lett. {\bf 87}, 011601 (2001),   
hep-ph/0012142; G.~Felder, L.~Kofman and A.~Linde, ``Tachyonic Instability   
and Dynamics of Spontaneous Symmetry Breaking,'' hep-th/0106179.   
  


\bibitem{OvrStein} 
B.~A.~Ovrut and P.~J.~Steinhardt,
``Supersymmetry And Inflation: A New Approach,''
Phys.\ Lett.\ B {\bf 133}, 161 (1983); B.~A.~Ovrut and P.~J.~Steinhardt,
``Inflationary Cosmology And The Mass Hierarchy In Locally Supersymmetric Theories,''
Phys.\ Rev.\ Lett.\  {\bf 53}, 732 (1984);
B.~A.~Ovrut and P.~J.~Steinhardt,
``Locally Supersymmetric Cosmology And The Gauge Hierarchy,''
Phys.\ Rev.\ D {\bf 30}, 2061 (1984);
B.~A.~Ovrut and P.~J.~Steinhardt,
``Supersymmetric Inflation, Baryon Asymmetry And The Gravitino Problem,''
Phys.\ Lett.\ B {\bf 147}, 263 (1984).

\bibitem{Eternal} 
A.~D.~Linde,
``Eternally Existing Selfreproducing Chaotic Inflationary Universe,''
Phys.\ Lett.\ B {\bf 175}, 395 (1986).





\bibitem{Hybrid}
A.~D.~Linde,
``Axions in inflationary cosmology,''
Phys.\ Lett.\ B {\bf 259}, 38 (1991);
A.~D.~Linde,
``Hybrid inflation,''
Phys.\ Rev.\ D {\bf 49}, 748 (1994)
[astro-ph/9307002].


 
\bibitem{F}
E.~J.~Copeland, A.~R.~Liddle, D.~H.~Lyth, E.~D.~Stewart and D.~Wands,
``False vacuum inflation with Einstein gravity,''
Phys.\ Rev.\ D {\bf 49}, 6410 (1994)
[astro-ph/9401011];
G.~R.~Dvali, Q.~Shafi and R.~Schaefer,
``Large scale structure and supersymmetric inflation without fine tuning,''
Phys.\ Rev.\ Lett.\  {\bf 73}, 1886 (1994)
[hep-ph/9406319];
E.~D.~Stewart,
``Inflation, supergravity and superstrings,''
Phys.\ Rev.\ D {\bf 51}, 6847 (1995)
[hep-ph/9405389].


\bibitem{D}
P.~Binetruy and G.~Dvali,
``D-term inflation,''
Phys.\ Lett.\ B {\bf 388}, 241 (1996)
[hep-ph/9606342];
E.~Halyo,
``Hybrid inflation from supergravity D-terms,''
Phys.\ Lett.\ B {\bf 387}, 43 (1996)
[hep-ph/9606423].


\bibitem{LythRiotto}
D.~H.~Lyth and A.~Riotto,
``Particle physics models of inflation and the cosmological density  perturbation,''
Phys.\ Rept.\  {\bf 314}, 1 (1999)
[hep-ph/9807278].
 

\bibitem{renata}
R.~Kallosh,
``N = 2 supersymmetry and de Sitter space,''
arXiv:hep-th/0109168; C.~Herdeiro, S.~Hirano and R.~Kallosh,
``String theory and hybrid inflation/acceleration,''
JHEP {\bf 0112} (2001) 027 [arXiv:hep-th/0110271]; K.~Dasgupta, C.~Herdeiro, S.~Hirano and R.~Kallosh,
``D3/D7 inflationary model and M-theory,''
arXiv:hep-th/0203019.
 

\bibitem{Vilenkin:xq}
P.~J.~Steinhardt,
``Natural Inflation,''
UPR-0198T
{\it Invited talk given at Nuffield Workshop on the Very Early Universe, Cambridge, England, Jun 21 - Jul 9, 1982};
A.~D.~Linde,
``Nonsingular Regenerating Inflationary Universe,''
  Cambridge Univ. preprint Print-82-0554, (1982);
A.~Vilenkin,
``The Birth Of Inflationary Universes,''
Phys.\ Rev.\ D {\bf 27}, 2848 (1983).
 


\bibitem{LLM}
A.~D.~Linde, D.~A.~Linde and A.~Mezhlumian,
``From the Big Bang theory to the theory of a stationary universe,''
Phys.\ Rev.\ D {\bf 49}, 1783 (1994)
[arXiv:gr-qc/9306035].



\bibitem{GuthVil}
A.~Borde, A.~H.~Guth and A.~Vilenkin,
``Inflation is not past-eternal,''
arXiv:gr-qc/0110012.
 
\bibitem{Bond} J. L. Sievers, J. R. Bond, J. K. Cartwright, C. R. Contaldi, B. S. Mason, S. T. Myers, S. Padin, T. J. Pearson, U.-L. Pen, D. Pogosyan, S. Prunet, A. C. S. Readhead, M. C. Shepherd, P. S. Udomprasert, L. Bronfman, W. L. Holzapfel, J. May (U. de Chile), ``Cosmological Parameters from Cosmic Background Imager Observations and Comparisons with BOOMERANG, DASI, and MAXIMA,'' astro-ph/0205387.


\bibitem{SperTur} D. Spergel and N. Turok, ``Textures and cosmic structure,'' Scientific
American {\bf 266}, 52 (1992).


 
\bibitem{KOST}   
J.~Khoury, B.~A.~Ovrut, P.~J.~Steinhardt and N.~Turok,
``The ekpyrotic universe: Colliding branes and the origin of the hot big  bang,''
Phys.\ Rev.\ D {\bf 64}, 123522 (2001)
[arXiv:hep-th/0103239].

  

\bibitem{cyclic} 
P.~J.~Steinhardt and N.~Turok,
``A cyclic model of the universe,''
arXiv:hep-th/0111030; P.~J.~Steinhardt and N.~Turok,
``Cosmic evolution in a cyclic universe,''
arXiv:hep-th/0111098;
P.~J.~Steinhardt and N.~Turok,
``Is vacuum decay significant in ekpyrotic and cyclic models?,''
arXiv:astro-ph/0112537;
 



   

   
   

\bibitem{KKL}   
R.~Kallosh, L.~Kofman and A.~D.~Linde,
``Pyrotechnic universe,''
Phys.\ Rev.\ D {\bf 64}, 123523 (2001)
[arXiv:hep-th/0104073].



\bibitem{KKLT}
R.~Kallosh, L.~Kofman, A.~D.~Linde and A.~A.~Tseytlin,
``BPS branes in cosmology,''
Phys.\ Rev.\ D {\bf 64}, 123524 (2001)
[arXiv:hep-th/0106241].



\bibitem{Khoury:2001bz}
J.~Khoury, B.~A.~Ovrut, N.~Seiberg, P.~J.~Steinhardt and N.~Turok,
``From big crunch to big bang,''
Phys.\ Rev.\ D {\bf 65}, 086007 (2002)
[arXiv:hep-th/0108187].


\bibitem{Khoury:2001zk}
J.~Khoury, B.~A.~Ovrut, P.~J.~Steinhardt and N.~Turok,
``Density perturbations in the ekpyrotic scenario,''
arXiv:hep-th/0109050.





    
   
\bibitem{Lyth:2001pf}
D.~H.~Lyth,
``The primordial curvature perturbation in the ekpyrotic universe,''
Phys.\ Lett.\ B {\bf 524}, 1 (2002)
[arXiv:hep-ph/0106153]; D.~H.~Lyth,
``The failure of cosmological perturbation theory in the new ekpyrotic  scenario,''
Phys.\ Lett.\ B {\bf 526}, 173 (2002)
[arXiv:hep-ph/0110007].


\bibitem{Brandenberger:2001bs}
R.~Brandenberger and F.~Finelli,
``On the spectrum of fluctuations in an effective field theory of the  ekpyrotic universe,''
JHEP {\bf 0111}, 056 (2001)
[arXiv:hep-th/0109004]; F.~Finelli and R.~Brandenberger,
``On the generation of a scale-invariant spectrum of adiabatic  fluctuations in cosmological models with a contracting phase,''
arXiv:hep-th/0112249.






\bibitem{Hwang:2002ks}
J.~c.~Hwang,
``Cosmological structure problem in the ekpyrotic scenario,''
Phys.\ Rev.\ D {\bf 65}, 063514 (2002)
[arXiv:astro-ph/0109045]; J.~c.~Hwang and H.~Noh,
``Non-singular big-bounces and evolution of linear fluctuations,''
arXiv:astro-ph/0112079; J.~Hwang and H.~Noh,
``Identification of perturbation modes and controversies in ekpyrotic  perturbations,''
arXiv:hep-th/0203193.


 
\bibitem{Martin:2001ue}
J.~Martin, P.~Peter, N.~Pinto Neto and D.~J.~Schwarz,
``Passing through the bounce in the ekpyrotic models,''
arXiv:hep-th/0112128; J.~Martin, P.~Peter, N.~Pinto-Neto and D.~J.~Schwarz,
``Comment on 'Density perturbations in the ekpyrotic scenario',''
arXiv:hep-th/0204222;  P.~Peter, J.~Martin, N.~Pinto-Neto and D.~J.~Schwarz,
``Perturbations in the ekpyrotic scenarios,''
arXiv:hep-th/0204227.


\bibitem{Durrer:2002jn}
R.~Durrer and F.~Vernizzi,
``Adiabatic perturbations in pre big bang models: Matching conditions and  scale invariance,''
arXiv:hep-ph/0203275.





\bibitem{HoravaWitten}   
P.~Ho\u{r}ava and E.~Witten,  Nucl.\ Phys.\  {\bf B475} (1996) 94   
[hep-th/9603142];  E.~Witten, Nucl.\ Phys.\ B {\bf 471}, 135 (1996)   
[hep-th/9602070]; P.~Ho\u{r}ava, Phys.\ Rev.\ D {\bf 54}, 7561 (1996)   
[hep-th/9608019];   
 T.~Banks and M.~Dine, Nucl.\ Phys.\ B {\bf 479}, 173 (1996) [hep-th/9605136];   
I.~Antoniadis and M.~Quiros,  
Phys.\ Lett.\ B {\bf 392}, 61 (1997) [hep-th/9609209]; H.~P.~Nilles,   
M.~Olechowski and M.~Yamaguchi, Phys.\ Lett.\ B {\bf 415}, 24 (1997) [hep-th/9707143];   
H.~P.~Nilles, M.~Olechowski and M.~Yamaguchi,Nucl.\ Phys.\ B {\bf 530} (1998) 43 [hep-th/9801030];   
H.~P.~Nilles, 
hep-ph/0004064.  

\bibitem{Benakli:1999sy}   
K.~Benakli, ``Scales and cosmological applications of M-theory,'' Phys.\   
Lett.\ B {\bf 447}, 51 (1999) [hep-th/9805181]; Z.~Lalak, S.~Pokorski and   
S.~Thomas, ``Beyond the standard embedding in M-theory on $S^1/Z_2$,''   
Nucl.\ Phys.\ B {\bf 549}, 63 (1999) [hep-ph/9807503].  

\bibitem{Donagi:2001fs}
R.~Y.~Donagi, J.~Khoury, B.~A.~Ovrut, P.~J.~Steinhardt and N.~Turok,
``Visible branes with negative tension in heterotic M-theory,''
JHEP {\bf 0111}, 041 (2001)
[arXiv:hep-th/0105199].


    
\bibitem{Moore:2000fs}   
G.~W.~Moore, G.~Peradze and N.~Saulina,
``Instabilities in heterotic M-theory induced by open membrane  instantons,''
Nucl.\ Phys.\ B {\bf 607}, 117 (2001)
[arXiv:hep-th/0012104].

   

\bibitem{universe} A.~Lukas, B.~A.~Ovrut and D.~Waldram,   
Nucl.\ Phys.\ B {\bf 532}, 43 (1998) [hep-th/9710208];   
Phys.\ Rev.\ D {\bf   
57}, 7529 (1998) [hep-th/9711197];
A.~Lukas, B.~A.~Ovrut, K.~S.~Stelle and D.~Waldram,    
Phys.\ Rev.\ D {\bf 59}, 086001 (1999) [hep-th/9803235]; A.~Lukas,   
B.~A.~Ovrut, K.~S.~Stelle and D.~Waldram, Nucl.\ Phys.\  {\bf B552}, 246 (1999) [hep-th/9806051].  



\bibitem{SeibergMoore}
H.~Liu, G.~Moore and N.~Seiberg,
``Strings in a time-dependent orbifold,''
arXiv:hep-th/0204168.


   

\bibitem{Nekrasov}
N.~A.~Nekrasov,
``Milne universe, tachyons, and quantum group,''
arXiv:hep-th/0203112.
 




\bibitem{Seiberg} N.~Seiberg, ``String Theory on a Time Dependent Orbifold,'' a seminar at the UCSC, 29 April 2002;
J. Polchinski, privite communication.



\bibitem{Rasanen:2001hf}
S.~Rasanen,
``On ekpyrotic brane collisions,''
Nucl.\ Phys.\ B {\bf 626}, 183 (2002)
[arXiv:hep-th/0111279].
 

\bibitem{negpot}
G.~Felder, A.~Frolov, L.~Kofman and A.~Linde,
``Cosmology with negative potentials,''
arXiv:hep-th/0202017; to be published in Phys. Rev. D
 


\bibitem{Banks:1996rr}
T.~Banks and M.~Dine,
``Phenomenology of strongly coupled heterotic string theory,''
arXiv:hep-th/9609046.




\bibitem{Creation}  
A.~D.~Linde,
``Quantum Creation Of The Inflationary Universe,''
Lett.\ Nuovo Cim.\  {\bf 39}, 401 (1984);
A.~Vilenkin,
``Quantum Creation Of Universes,''
Phys.\ Rev.\ D {\bf 30}, 509 (1984).
 



\bibitem{Kallosh:2000ve}
R.~Kallosh, L.~Kofman, A.~D.~Linde and A.~Van Proeyen,
``Superconformal symmetry, supergravity and cosmology,''
Class.\ Quant.\ Grav.\  {\bf 17}, 4269 (2000)
[arXiv:hep-th/0006179].
 


\bibitem{ford} L.~H.~Ford,
 ``Gravitational Particle Creation And Inflation,''
Phys.\ Rev.\ D {\bf 35}, 2955 (1987).

 

\bibitem{grib} Ya. B. Zel'dovich
and A. A. Starobinsky,  ``Particle Production And Vacuum Polarization In An Anisotropic Gravitational Field,'' Zh. Exp. Theor. Fiz. {\bf 61}, 2161 (1971)
[Sov. Phys. JETP {\bf 34}, 1159 (1972)]; L.~P.~Grishchuk,
 ``The Amplification Of Gravitational Waves And Creation Of Gravitons In The Isotropic Universes,''
Lett.\ Nuovo Cim.\  {\bf 12}, 60 (1975); Sov. Phys. JETP  {\bf 40}, 409
(1975); S.~G.~Mamaev, V.~M.~Mostepanenko and A.~A.~Starobinsky,
 ``Particle Creation From Vacuum Near An Homogeneous Isotropic Singularity,''
Zh.\ Eksp.\ Teor.\ Fiz.\  {\bf 70}, 1577 (1976) [Sov. Phys. JETP {\bf 43}, 823
(1976)];
 A.~A.~Grib, S.~G.~Mamaev and V.~M.~Mostepanenko,
 ``Particle Creation From Vacuum In Homogeneous Isotropic Models Of The Universe,''
Gen.\ Rel.\ Grav.\  {\bf 7}, 535 (1976); A.A. Grib, S.G. Mamaev, and V. M. Mostepanenko, {\it
Vacuum Quantum Effects in Strong Fields} (Energoatonizdat, Moscow,
1988).

 
\bibitem{Toporensky:1999pk}
A.~V.~Toporensky,
``The degree of generality of inflation in FRW models with massive scalar  field and hydrodynamical matter,''
Grav.\ Cosmol.\  {\bf 5}, 40 (1999)
[arXiv:gr-qc/9901083].
 

\bibitem{Tolley:2002cv}
A.~J.~Tolley and N.~Turok,
``Quantum fields in a big crunch/big bang spacetime,''
arXiv:hep-th/0204091.
 




\bibitem{Tolman-1931}
R.C. Tolman, Phys. Rev. {\bf 38}, 1758 (1931);
G. Lema\^itre, Ann. Soc. Sci. Bruxelles A {\bf 53}, 51 (1933), translated in Gen Rel. Grav. {\bf 29}, 935 (1997);
 R. Dicke, and P.J.E. Peebles, in: {\it General relativity},
eds., S.W. Hawking, and W. Israel,
(Cambridge Univ. Press, Cambridge, 1979), 504p.

\bibitem{Tolman-1934}
 R.C. Tolman, {\it Relativity, thermodynamics and cosmology}, 
(Oxford Univ. Press, London, 1934).

\bibitem{Peebles-1993}
P.J.E. Peebles, {\it Principles of physical cosmology}
(Princeton Univ. Press, Princeton, 1993).


\bibitem{Markov}
M.~A.~Markov,
``Problems Of A Perpetually Oscillating Universe,''
Annals Phys.\  {\bf 155}, 333 (1984).

\bibitem{LindeReview}
A.~D.~Linde,
``The Inflationary Universe,''
Rept.\ Prog.\ Phys.\  {\bf 47}, 925 (1984), Appendix 1.

\bibitem{BrVafa}
R.~H.~Brandenberger and C.~Vafa,
``Superstrings In The Early Universe,''
Nucl.\ Phys.\ B {\bf 316}, 391 (1989).

\bibitem{Brandenberger:1993ef}
R.~H.~Brandenberger, V.~Mukhanov and A.~Sornborger,
``A Cosmological theory without singularities,''
Phys.\ Rev.\ D {\bf 48}, 1629 (1993)
[arXiv:gr-qc/9303001].

\bibitem{PBB} G.~Veneziano,
``Scale factor duality for classical and quantum strings,''
Phys.\ Lett.\ B {\bf 265}, 287 (1991); M.~Gasperini and G.~Veneziano,
``Pre - big bang in string cosmology,''
Astropart.\ Phys.\  {\bf 1}, 317 (1993)
[arXiv:hep-th/9211021].


\bibitem{Kaloper} 
M.~S.~Turner and E.~J.~Weinberg,
``Pre-big-bang inflation requires fine tuning,''
Phys.\ Rev.\ D {\bf 56}, 4604 (1997)
[arXiv:hep-th/9705035];
N.~Kaloper, A.~D.~Linde and R.~Bousso,
``Pre-big-bang requires the universe to be exponentially large from the  very beginning,''
Phys.\ Rev.\ D {\bf 59}, 043508 (1999)
[arXiv:hep-th/9801073];
A.~Buonanno and T.~Damour,
``The fate of classical tensor inhomogeneities in pre-big-bang string  cosmology,''
Phys.\ Rev.\ D {\bf 64}, 043501 (2001)
[arXiv:gr-qc/0102102].



\bibitem{QuintInfl} B.Spokoiny,
``Deflationary universe scenario,''
Phys.\ Lett.\ B {\bf 315}, 40 (1993)
[arXiv:gr-qc/9306008];
MM.~Joyce,
``On the expansion rate of the universe at the electroweak scale,''
Phys.\ Rev.\ D {\bf 55}, 1875 (1997)
[arXiv:hep-ph/9606223]; M.~Joyce and T.~Prokopec,
``Turning around the sphaleron bound: Electroweak baryogenesis in an  alternative post-inflationary cosmology,''
Phys.\ Rev.\ D {\bf 57}, 6022 (1998)
[arXiv:hep-ph/9709320];
P.~J.~Peebles and A.~Vilenkin,
``Quintessential inflation,''
Phys.\ Rev.\ D {\bf 59}, 063505 (1999)
[arXiv:astro-ph/9810509];  D.~J.~Chung, E.~W.~Kolb and A.~Riotto,
 ``Superheavy dark matter,''
Phys.\ Rev.\ D {\bf 59}, 023501 (1999)
[arXiv:hep-ph/9802238]; V.~Kuzmin and I.~Tkachev,
``Matter creation via vacuum fluctuations in the early universe and  observed ultra-high energy cosmic ray events,''
Phys.\ Rev.\ D {\bf 59}, 123006 (1999)
[arXiv:hep-ph/9809547];
G.~N.~Felder, L.~Kofman and A.~D.~Linde,
``Gravitational particle production and the moduli problem,''
JHEP {\bf 0002}, 027 (2000)
[arXiv:hep-ph/9909508];
G.~F.~Giudice, A.~Riotto and I.~I.~Tkachev,
``The cosmological moduli problem and preheating,''
JHEP {\bf 0106}, 020 (2001)
[arXiv:hep-ph/0103248].


\bibitem{Instant}
G.~N.~Felder, L.~Kofman and A.~D.~Linde,
``Instant preheating,''
Phys.\ Rev.\ D {\bf 59}, 123523 (1999)
[arXiv:hep-ph/9812289].

\bibitem{NO}
G.~N.~Felder, L.~Kofman and A.~D.~Linde,
``Inflation and preheating in NO models,''
Phys.\ Rev.\ D {\bf 60}, 103505 (1999)
[arXiv:hep-ph/9903350].

 

\bibitem{cyclicnew} P.~J.~Steinhardt and N.~Turok, ``A Cyclic Model of the Universe,''  Science,    Published online April 25, 2002; 10.1126/science.1070462 (Science Express Research Articles).

\bibitem{cyclicastro} P.~J.~Steinhardt and N.~Turok,
``The cyclic universe: An informal introduction,''
arXiv:astro-ph/0204479.

\bibitem{Hollands:2002yb}
S.~Hollands and R.~M.~Wald,
``An alternative to inflation,''
arXiv:gr-qc/0205058.

\bibitem{alt} L. Kofman, A. Linde and V. Mukhanov, in preparation.

\end{thebibliography}
\end{document}